\documentclass[journal,twoside,10pt,twocolumn]{IEEEtran}
\usepackage{graphicx,amssymb,amsmath,amsthm,psfrag,stfloats,color,dsfont,subfigure,listings,epstopdf,cite,bm}
\usepackage{algorithm,algorithmic}
\usepackage[skip=-3pt]{caption}

\begin{document}

\title{\fontsize{19pt}{19}\selectfont Simultaneous Unicast and Multicast Transmissions in Stacked Intelligent Metasurfaces-assisted HAPS Wireless Networks: Performance Analysis and Optimization}

\author{Ngoc Phuc Le, \IEEEmembership{Member, IEEE}, and Mohamed-Slim Alouini, \IEEEmembership{Fellow, IEEE}
\thanks{The authors are with the Division of Computer, Electrical and Mathematical Sciences and Engineering, King Abdullah University of Science and Technology, Thuwal 23955-6900, Saudi Arabia (e-mails: phucle.ngoc@kaust.edu.sa;slim.alouini@kaust.edu.sa).}

\vspace*{-0.5cm}}
\maketitle

\begin{abstract}
In this paper, we investigate high-altitude platform station (HAPS) wireless networks for simultaneous non-orthogonal unicast and multicast transmissions. Specifically, stacked intelligent metasurface (SIM)-based wave-domain beamforming is proposed to enable efficient HAPS-to-ground communications. Also, the system performance is investigated from an energy-efficiency (EE) perspective, which is a crucial for HAPS operations. For performance analysis, we derive approximate closed-form expressions for the outage probability over Rician fading channels. For EE optimization, we jointly optimize the transmit power and the SIM phase-shifts for the maximal EE. Two methods are proposed to solve this non-convex optimization problem. The first method develops an efficient alternating optimization (AO) framework based on golden-section search and projected gradient ascent (PGA) for transmit power and phase-shift optimization, respectively. The second method uses unsupervised deep neural network (DNN) that does not require labeling. Performance comparison between the two methods, as well as with other benchmarks schemes are examined. Additionally, the impacts of the number of SIM elements per layers, the number of SIM layers, the maximum transmit power on the EE performance are evaluated. Simulation results are provided to demonstrate the performance of the proposed systems.
\end{abstract}

\begin{IEEEkeywords}
\noindent High-Altitude Platform Station (HAPS), Stacked Intelligent Metasurfaces (SIM), Alternative Optimization (AO), Machine Learning (ML), Energy Efficiency (EE).
\end{IEEEkeywords}

\makeatletter
\long\def\@makecaption#1#2{\ifx\@captype\@IEEEtablestring%
\footnotesize\begin{center}{\normalfont\footnotesize #1}\\
{\normalfont\footnotesize\scshape #2}\end{center}%
\@IEEEtablecaptionsepspace
\else
\@IEEEfigurecaptionsepspace
\setbox\@tempboxa\hbox{\normalfont\footnotesize {#1.}~~ #2}%
\ifdim \wd\@tempboxa >\hsize%
\setbox\@tempboxa\hbox{\normalfont\footnotesize {#1.}~~ }%
\parbox[t]{\hsize}{\normalfont\footnotesize \noindent\unhbox\@tempboxa#2}%
\else
\hbox to\hsize{\normalfont\footnotesize\hfil\box\@tempboxa\hfil}\fi\fi}
\makeatother

\section{Introduction}
\IEEEPARstart{T}he Recommendation defining frameworks and objectives of the future developments for International Mobile Telecommunications-2030 and Beyond (IMT-2030) was recently released in 2023 \cite{IMT}. It outlines usage scenarios, capabilities, and the roadmap for the sixth generation of mobile communication technology (6G). While the requirements and evaluation methodology phase is still ongoing, core technologies have been identified, including extreme MIMO (multiple input-multiple output), reconfigurable intelligent surfaces (RIS), integrated sensing and communications (ISAC), digital twins, and non-terrestrial networks. These technologies are expected to meet the stringent requirements and diverse application scenarios anticipated in 6G networks \cite{Dan}.

Over the last few years, RIS has emerged as a promising paradigm for enhancing wireless communication performance \cite{Di}-\hspace{1pt}\cite{Wu}. RIS enables customized radio wave propagation by intelligently controlling the phase and amplitude of its elements, thereby improving link reliability, spectral efficiency, and network coverage \cite{Di}. Recent advances and the road to 6G for this revolutionary technology was discussed in \cite{Wu}. Meanwhile, non-terrestrial networks (NTN) was first officially introduced in 3GPP-Release 17 (3rd Generation Partnership Project) in 2022 \cite{GPP}. It involves communication nodes located on satellites, high altitude platform stations (HAPS), or unmanned aerial vehicles (UAV). NTN is envisioned to extend coverage to remote, rural, and underserved areas, provide resilient communications, and support massive IoT networks \cite{Aza}. Consequently, it is essential to study the integration of RIS and NTN networks to exploit the potential benefits of these technologies for 6G networks. In fact, RIS-assisted NTN networks have attracted considerable interests from the research community \cite{Ye}.

\subsection{Review of Related Works} 
\textit{\textbf{HAPS with RIS}}: 
HAPS is an aerial platform that operates in the stratosphere, i.e., between 17 and 22 kilometers in altitude. It offers several distinctive advantages that make it highly attractive for 6G NTN networks, including a balance between wide-area coverage and low latency, along with flexible and rapid deployment capabilities \cite{Kar}. Therefore, many research works have considered RIS for HAPS networks \cite{Alf}-\hspace{1pt}\cite{Gao1}. In \cite{Alf}, RIS-HAPS was proposed to support edge users where terrestrial networks is a cost-ineffective approach. The authors focused on a resource-efficient optimization problem that maximizes the number of connected users, while minimizing the total power consumption. In \cite{Ji}, a deployment of active RIS for HAPS was proposed, where a joint optimization of power allocation and active reflecting parameters for maximizing the upper bound of the spectral efficiency was formulated and solved. Additionally, performance analysis of RIS-HAPS systems in terms of outage probability (OP), average symbol error rate and ergodic capacity was carried out in \cite{Sha}. It is worth noting that these studies assume stationary HAPS platforms. In contrast, aerodynamic RIS-assisted HAPS system is considered in \cite{Azi}, where the authors derived a closed-form solution for the RIS phase shifts to simultaneously maximize the channel gain and minimize the delay spread upper bound. Also, the joint optimization of the HAPS trajectory and RIS phase shifts for maximizing the received signal-to-noise ratio (SNR) at ground users under an unknown and dynamic radio environment was investigated in \cite{Gao1}. 

\textit{\textbf{SIM-based Wireless Systems}}: 
Conventional RIS-based systems are typically based on single-layer meta-surface for reflection or transmission, which offers limited signal processing capabilities and performance enhancements. Recently, a novel scheme of stacked-intelligent metasurface (SIM), which composes of cascaded metasurfaces, was proposed by J. An \textit{et al.} \cite{An}-\hspace{1pt}\cite{An1}. With the SIM scheme, complex signal processing can be performed directly in the electromagnetic wave domain by optimizing the phase-shifts of meta-atoms. Thus, SIM reduces processing delays, alleviates the need for high-resolution digital-to-analog converters/analog-to-digital converters (DACs/ADCs), and requires fewer radio frequency (RF) chains, which results in lower hardware costs and energy power consumption \cite{An2}. Motivated by this, several research works have explored the potential benefits of SIM from various perspectives and under diverse system models \cite{An}-\hspace{1pt}\cite{Hua}.
     
Most available research works have focused on power allocation and wave-based beamforming to improve the system performance \cite{An}-\hspace{1pt}\cite{Jia}. For maximizing achievable sum-rates, \cite{An} considers multi-user multiple-input single output (MISO) systems, while \cite{An1} focuses on single-user MIMO systems. Also, maximizing the sum-rates of the multi-user MISO systems based on statistical channel state information (CSI) was investigated in \cite{Pap1}. In these studies, alternating optimization (AO) methods were developed to solve the joint optimizations, where the optimal phase-shifts are obtained by using the gradient ascent algorithms. On the other hand, a deep reinforcement learning approach (DRL) was proposed in \cite{Liu} for the sum-rate maximization. For an energy-efficiency (EE) perspective, optimizing phase-shifts for energy-efficient SIM-based MIMO systems was reported in \cite{Per}. Also, \cite{Niu7} proposed incorporating meta-fibers into SIM structures to reduce the number of layers and improving the EE. Furthermore, channel estimation in SIM-based systems was considered in \cite{Yao} and \cite{Pap2} for Rayleigh fading and Rician fading channels, respectively. Also, a hardware platform of SIM using one-bit unit cells were developed in \cite{Jia}, where the authors evaluated and validated the performance of SIM-based systems via experiments.

\textit{\textbf{Integration of SIM with Emerging Technologies}}: 
The integration of SIM with emerging wireless technologies has been investigated in \cite{Hu}-\cite{Hua}. In particular, in \cite{Hu} and \cite{Shi}, SIM is incorporated into cell-free massive MIMO systems for enhanced sum-rates. For the downlink, the authors of \cite{Hu} jointly optimized the transmit power allocation and the SIM phase shifts for maximized sum rate by using an AO approach. For the uplink, the authors of \cite{Shi} derived closed-form expressions of the spectral efficiency and devised an iterative algorithm based on statistical CSI for maximized sum rate. In \cite{Li}, SIM-based systems is considered for near-field communications, where optimized phase shifts are obtained based on a proposed layer-by-layer iterative algorithm. Meanwhile, SIM-based LEO satellite communications using statistical CSI was studied in \cite{Lin}, where the authors jointly optimize power allocation and SIM phase shift for maximizing the sum-rate. Furthermore, motivated by the potential of SIM for implementing beamforming and direction of arrival (DoA) estimation, some works have studied SIM-based integrated sensing and communication, e.g., \cite{Niu1}-\hspace{1pt}\cite{Lis}. Also, the wave-based computing paradigm enabled by SIM was shown to be effective for implementing image recognition in semantic communications \cite{Hua}.

\subsection{Open Research Problem and Contributions}
The integration of HAPS into 6G architectures offers a compelling solution to enable seamless coverage in underserved, remote, and mobility-challenged environments. To further enhance the performance of HAPS-based networks, we propose to incorporate SIM into HAPS systems in this study. The key motivation for using SIM rather than a single-layer RIS on HAPS is that the SIM architecture can achieve better beamforming flexibility and energy efficiency. This reduces the required transmit power and/or surface aperture, which is critical for HAPS platforms. Also, it is worth noting that frameworks to support multicast and broadcast services in NTN environments will be included in 3GPP Release 19 \cite{GPP1}. Motivated by this, we consider non-orthogonal unicast and multicast transmissions in the SIM-HAPS system model. The main contributions of this work are summarized below.
\begin{itemize}
	\item The SIM-HAPS system model is proposed that could exploit the potential benefits of SIM-based beamforming for efficient data transmissions from the HAPS to ground stations. In this model, a hybrid digital precoding and SIM-based beamforming is adopted to mitigate interference between unicast and multicast signals as well as among unicast signals. As a result, the system performance is improved in terms of outage probability and energy efficiency. 
	\item Performance of the SIM-HAPS system, where a digital precoder is designed based on a MRT/ZF (maximum ratio transmission/zero-forcing) method, is analyzed. In particular, we derive approximate closed-form expressions of the outage probability over the correlated Rician fading channel based on gamma approximation and saddle point approximation (SPA) techniques. The highly accuracy of these expressions is validated through simulations.
	\item The key system parameters, including the transmit power of HAPS and phase-shifts of the SIM, are optimized for the maximal EE. Specifically, we propose two methods to solve the non-convex optimization problem, namely alternative optimization (AO) optimization and unsupervised deep neural network (DNN)-based approaches. The AO approach deploys one-dimensional search to find the level of transmit power, and uses projected gradient ascent (PGA) for phase-shift updates. Meanwhile, the unsupervised DNN-based approach optimizes system parameters by updating the neural network weights without requiring labeled data.
	\item The achieved EE performance in SIM-HAPS systems under different scenarios are compared, including: AO versus unsupervised DNN, joint optimization of transmit power and SIM phase-shifts versus individual optimization of each component. We demonstrate that the system with the joint optimization solutions achieve higher EE than those with only transmit power optimization or phase-shift optimization.		
	\item The impacts of the transmit power and the SIM parameters, such as a number of SIM layers and a size of each layer, on the EE performance are evaluated. Our results indicate that an appropriate selection of the number of SIM layers and the maximum transmit power is essential to attain maximal EE.
\end{itemize}

\subsection{Organization of the Paper}
The remaining of this paper is organized as follows. In Section II, we describe the proposed SIM-HAPS system model. Section III derives the OP expressions, while Section IV optimizes the EE. In particular, Section IV.B proposes the AO approach to solve the EE maximization problem, whereas Section IV.C develops an unsupervised DNN-based solution. Section V provides simulation results and discussions. Finally, Section VI presents the conclusions of this paper.

\section{SIM-HAPS Wireless System Model}
We consider a HAPS-based wireless system model that consists of one HAPS and $K$ single-antenna ground stations (GSs) as shown in Fig. 1. The HAPS is equipped with $M$ antennas ($M \geq K+1$) and a SIM device, which enables transmit beamforming in the wave-domain\footnote{Each SIM layer is an ultra-thin, lightweight passive planar metasurface, and the inter-layer spacing in SIM is very small. Thus, the additional weight introduced by stacking multiple SIM layers remains modest. Moreover, the stacked architecture does not significantly increase the effective area exposed to stratospheric winds compared to single-layer RIS.}. In this system, the HAPS transmits both unicast and multicast signals in the same resource blocks to the GSs. The multicast signal delivers shared information to all GSs, while the unicast signals are dedicated transmissions intended for individual GSs. We note that the main objective of this work is to investigate the fundamental performance gains and EE enabled by SIM architectures. Thus, we assume perfectly aligned stacked metasurfaces, stable HAPS position, and perfect CSI to facilitate tractable analysis\footnote{Several practical issues, including HAPS platform dynamics, misalignment modeling, signaling overhead as well as the time/bandwidth consumed by pilot training associated with the estimated channel phase, are important topics that are left for future investigation.}. This is similar to prior SIM-based system studies in the literature, e.g., \cite{Alf}-\hspace{1pt}\cite{An2}, \cite{Liu}-\hspace{1pt}\cite{Per}.

\begin{figure}[t] 
	\centering{\includegraphics[width=0.45\textwidth]{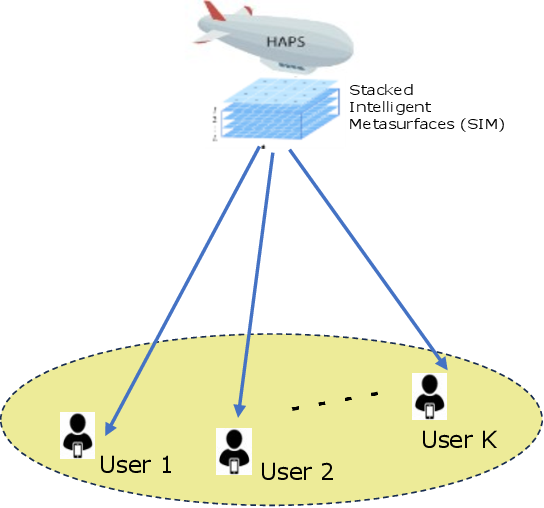}}
	\caption{A SIM-HAPS wireless system model.}
	\label{Fig1}
\end{figure}

\subsection{SIM Model}
Let us first describe the model of the SIM mounted on HAPS. In our system model, the SIM is composed of $L$ meta-surfaces layers, each consists of $N$ meta-atoms (i.e., SIM elements). Let $\mathcal{L} = \{1,2,...,L\}$, $\mathcal{N} =\{1,2,...,N\}$, and $\mathcal{K} = \{1,2,...,K\}$ denote the set of SIM layers, the SIM elements on each layer, and the ground users, respectively. Transmit beamforming in the wave domain is realized by properly adjusting phase-shifts of SIM elements via a smart controller attached to SIM. The phase-shift of the $n$-th element on the $l$-th layer is denoted as $\theta_n^l \in [0, 2\pi), \forall n\in \mathcal{N}, \forall l \in \mathcal{L}$. Then, the phase-shift matrix induced by the $l$-th SIM layer is given as $\mathbf{\Theta}_l=\operatorname{diag}(e^{j\theta_1^l}, e^{j\theta_2^l}, \cdots, e^{j\theta_N^l}) \in \mathbb{C}^{N\times N}, \forall l \in \mathcal{L}$. Also, let $\mathbf{\Psi}_l \in \mathbb{C}^{N\times N}, \forall l \in \mathcal{L} \setminus \{1\},$ denotes the propagation matrix from the $(l-1)$-th layer to the $l$-th layer. Each entry of this matrix can be modeled by the Rayleigh-Sommerfeld diffraction theory \cite{Lin}. In particular, we can express \cite{An}
\begin{align} \label{S1}
\psi_{n,n'}^l=\frac{d_xd_y\cos \chi_{n,n'}^l}{d_{n,n'}^l}\left(\frac{1}{2\pi d_{n,n'}^l}-j\frac{1}{\lambda}\right)e^{j2\pi d_{n,n'}^l/\lambda},
\end{align}        
\noindent where $\lambda$ is the wavelength, $d_{n,n'}^l$ is the propagation distance between the meta-atoms, $\chi_{n,n'}^l$ denotes the angle between the propagation direction and the normal direction of the $(l-1)$-th layer, and $d_x d_y$ denotes the size of each meta-atom. This model is also applied for the propagation from the transmit antennas to the first layer of SIM. Specifically, the propagation vector from the $m$-th antenna to the first layer is denoted as $\bm{\psi}_m^1 \in \mathbb{C}^{N\times 1}$, where the $n$-th entry of $\bm{\psi}_m^1$, (i.e., $\psi_{n,m}^1$) is evaluated from \eqref{S1}. Consequently, the beamforming matrix based on SIM can be expressed as
\begin{align} \label{S2}
\mathbf{B}=\mathbf{\Theta}_L\mathbf{\Psi}_L\mathbf{\Theta}_{L-1}\mathbf{\Psi}_{L-1} \cdots \mathbf{\Theta}_2\mathbf{\Psi}_2\mathbf{\Theta}_1 \in \mathbb{C}^{N\times N}.
\end{align}   

\subsection{Downlink Transmission}
At the HAPS, the multicast data intended for all GSs is encoded as the common stream $s_0$. Meanwhile, the private data intended for the $k$-th GS is encoded as the private stream $s_k, \forall k \in \mathcal{K}$. As a result, one common stream and $K$ private streams are transmitted simultaneously by the HAPS to the GSs. We consider a digital precoder to handle interference among data streams. Specifically, we can express the precoding vector as $\mathbf{w}_k=\sqrt{p_k}\mathbf{\bar{w}}_k \in \mathbb{C}^{M\times 1}, (k=0,1,...,K)$, where $p_k$ is the transmit power associated with the $k$-th stream, and $\mathbf{\bar{w}}_k$ is the beam direction. Thus, the received signal at the $k$-th GS is obtained as 
\begin{align} \label{S7}
y_k&= \mathbf{h}_k^H\mathbf{B}\boldsymbol{\Psi}_1\mathbf{\bar{w}}_0\sqrt{p_0}s_0+\sum_{r=1}^K\mathbf{h}_k^H\mathbf{B}\boldsymbol{\Psi}_1\mathbf{\bar{w}}_r\sqrt{p_r}s_r+n_k\nonumber\\ 
&=\mathbf{g}_k\mathbf{\bar{w}}_0\sqrt{p_0}s_0+\sum_{r=1}^K\mathbf{g}_k\mathbf{\bar{w}}_r\sqrt{p_r}s_r+n_k,
\end{align} 
\noindent where $\mathbf{h}_k^H \in \mathbb{C}^{1\times N}$ denotes the baseband equivalent channel from the last SIM layer to the $k$-th ground user, $\mathbb{E}\{|s_k|^2\}=1$, and $n_k \sim \mathcal{CN}(0,N_0)$ denotes the noise with the noise power $N_0$. Also, $\mathbf{\Psi}_1= [\bm{\psi}_1^1, \bm{\psi}_2^1, \cdots, \bm{\psi}_M^1] \in \mathbb{C}^{N\times M}$ is the propagation matrix from the antennas to the first layer of the SIM, and $\mathbf{g}_k \triangleq \mathbf{h}_k^H\mathbf{B}\boldsymbol{\Psi}_1 \in \mathbb{C}^{1\times M}$ is the equivalent channel from the antennas to the $k$-th GS. Additionally, it is worth noting that the total transmit power $P$ is constrained by $P=\sum_{k=0}^K p_k \leq P_{max}$, where $P_{max}$ is the maximum transmit power at the HAPS.

At each GS, the decoding process is based on a successive interference cancellation techniques (SIC). Similar to several works on non-orthogonal unicast and multicast transmissions, e.g., \cite{Kim}-\hspace{1pt}\cite{Liz}, we assume that the multicast data, which is intended for multiple users, should have a higher decoding priority. Accordingly, the $k$-th user first decodes the multicast signal $s_0$ by treating all private unicast signals as noise. Then, the multicast signal is removed from the total received signal, and the user will decode its private signal by treating all the other private signals as noise. Mathematically, we can express the signal-to-interference-plus-noise ratios (SINR) of the multicast and unicast stream, respectively, as
\begin{align} \label{S5}
\gamma_k^{\mathcal{M}}=\frac{|\mathbf{g}_k\mathbf{\bar{w}}_0|^2p_0}{\sum_{r=1}^K|\mathbf{g}_k\mathbf{\bar{w}}_r|^2p_r+N_0},
\end{align} 
\noindent and
\begin{align} \label{S6}
\gamma_k^{\mathcal{U}}=\frac{|\mathbf{g}_k\mathbf{\bar{w}}_k|^2p_k}{\sum_{r=1, r \neq k}^K|\mathbf{g}_k\mathbf{\bar{w}}_r|^2p_r+N_0}.
\end{align} 

In this work, we consider maximum ratio transmision (MRT) for the multicast stream, whereas the zero-forcing (ZF) is considered for the unicast streams. For convenience, let us define $\mathbf{G}=[\mathbf{g}_1^T, \mathbf{g}_2^T, \cdots, \mathbf{g}_K^T] \in \mathbb{C}^{M\times K}$ and $\mathbf{D}=\operatorname{diag}(||\mathbf{g}_1||, ||\mathbf{g}_2||,\cdots, ||\mathbf{g}_K||) \in \mathbb{C}^{K\times K}$ as the channel matrix and the diagonal matrix, respectively, where $||\cdot||$ denotes the Euclidean norm. Then, the beam direction vector $\mathbf{\bar{w}}_k, (k=1,2,..,K)$, is the column of the matrix $\mathbf{W}=\mathbf{G}^H(\mathbf{G}\mathbf{G}^H)^{-1}\mathbf{D}$. Also, the beam direction vector associated with the multicast stream is given as $\mathbf{\bar{w}}_0=\sum_{r=1}^K \mathbf{\bar{w}}_r$ \cite{Vu}. Note that with this design, we obtain $\mathbf{g}_k\mathbf{\bar{w}}_0=||\mathbf{g}_k||$, $\mathbf{g}_k\mathbf{\bar{w}}_k=||\mathbf{g}_k||$, and $\mathbf{g}_k\mathbf{\bar{w}}_r=0 (r \in \mathcal{K}, r \neq k)$. Consequently, the SINR of the common and private streams can be now expressed as (cf. \eqref{S5}-\eqref{S6})
\begin{align} \label{S9}
\gamma_k^{\mathcal{M}}=\frac{||\mathbf{g}_k||^2p_0}{||\mathbf{g}_k||^2p_k+N_0},
\end{align} 
\noindent and
\begin{align} \label{S10}
\gamma_k^{\mathcal{U}}=\frac{||\mathbf{g}_k||^2p_k}{N_0}.
\end{align} 
\noindent Based on this, the achievable rates of the private messages are obtained as 
\begin{align} \label{S11}
R_k=\log_2(1+\gamma_k^{\mathcal{U}}), \forall k \in \mathcal{K}. 
\end{align} 
\noindent Also, to guarantee that the multicast message is successfully decoded by all GSs, the achievable rate of the multicast message is calculated as 
\begin{align} \label{S12}
R_0=\min_{k=1,2,...,K}\{\log_2(1+\gamma_k^{\mathcal{M}})\}.
\end{align} 

\subsection{HAPS-to-Ground Users Channel Model}
For HAPS-to-ground channels, it is typically that the line-of-sight (LoS) path exists. Thus, we consider the Rician fading to model the channels. The channel from the last layer of SIM to the $k$-th user, denoted by $\mathbf{h}_k \in \mathbb{C}^{N\times1}$, is given as
\begin{align} \label{C1}
\mathbf{h}_k=\sqrt{\beta_k}\left(\sqrt{\frac{\kappa}{1+\kappa}}\mathbf{h}_{k,LoS}+\sqrt{\frac{1}{1+\kappa}}\mathbf{h}_{k,NLoS}\right),
\end{align} 
\noindent where $\beta_k$ is the path-loss coefficient, $\kappa$ is the Rician factor, $\mathbf{h}_{k,LoS} \in \mathbb{C}^{N\times1}$ is the LoS component, and $\mathbf{h}_{k,NLoS}\in \mathbb{C}^{N\times1}$ is the NLoS component. The distance-dependent large-scale path loss is calculated as $\beta_k=G_tG_r\left(\frac{\lambda}{4\pi d_k}\right)^2$, where $d_k$ is the distance from the HAPS to the $k$-th ground user. Also, $\mathbf{h}_{k,NLoS} \sim \mathcal{CN}(\mathbf{0},\mathbf{R})$, where $\mathbf{R} \in \mathbb{C}^{N\times N}$ is the covariance matrix that represents the spatial correlation between different meta-atoms. Each entry of this matrix is given as $\mathbf{R}_{n,n'} =\frac{\sin(2\pi d_{n,n'}/\lambda)}{2\pi d_{n,n'}/\lambda}$, where $d_{n,n'}$ denotes the distance between the meta-atoms \cite{An}, \cite{Pap2}. Meanwhile, the deterministic LoS component is computed as $\mathbf{h}_{k,LoS}=\mathbf{a}_k(\varphi_{k,x}, \varphi_{k,y})$, where the steering vector of the last layer is given as 
\begin{align} \label{C2}
&\mathbf{a}_k(\varphi_{k,x}, \varphi_{k,y}) = \nonumber \\
&[1,e^{j\frac{2\pi}{\lambda}d_e\cos \varphi_{k,x}\sin \varphi_{k,y}}, \cdots, e^{j\frac{2\pi}{\lambda}(N_x-1)d_e\cos \varphi_{k,x}\sin \varphi_{k,y}}]^T \otimes \nonumber \\
& [1,e^{j\frac{2\pi}{\lambda}d_e\sin \varphi_{k,x}\sin \varphi_{k,y}}, \cdots, e^{j\frac{2\pi}{\lambda}(N_y-1)d_e\sin \varphi_{k,x}\sin \varphi_{k,y}}]^T,
\end{align} 
\noindent where $\varphi_{k,x}$ and $\varphi_{k,y}$ represent the azimuth and the elevation angles of departure (AoD) at the HAPS to the $k$-th user, and $\otimes$ denotes the Kronecker product.

\section{Analysis of Outage Probability}
In this system, the outage event at the $k$-th user occurs when it fails to decode either $s_0$ or $s_k$. Thus, the outage probability (OP) can be expressed as
\begin{align} \label{O1}
P_{out,k}=1-Pr(\gamma_k^{\mathcal{M}}>\gamma_0^{th}, \gamma_k^{\mathcal{U}}>\gamma_k^{th}),
\end{align} 
\noindent where $\gamma_0^{th}$ and $\gamma_k^{th}$ are the SNR thresholds for the multicast signal and the unicast signal of the $k$-th user. Using the SINR values defined in \eqref{S9} and \eqref{S10}, we can rewrite the OP as
\begin{align} \label{O2}
P_{out,k}=Pr(||\mathbf{g}_k||^2<\xi_k),
\end{align} 
\noindent where $\xi_k \triangleq \max\left\{\frac{\gamma_0^{th}N_0}{p_0-\gamma_0^{th}p_k},\frac{\gamma_k^{th}N_0}{p_k}\right\}$ if $p_0>\gamma_0^{th}p_k$. By using gamma distribution to approximate statistical distribution of $||\mathbf{g}_k||^2$, we arrive at the following result.
 
\noindent\textbf{\textit{Theorem 1:}} \textit{The approximate expression of the OP for the $k$-th user is given as}
\begin{align} \label{O3}
P_{out,k}=\frac{1}{\Gamma(\vartheta_k)}\gamma(\vartheta_k,\xi_k/\varkappa_k), 
\end{align}
\textit{where $\vartheta_k = m_Z^2/v_Z$, $\varkappa_k=v_Z/m_Z$, and $m_Z$ and $v_Z$ are the mean and the variance of $Z =||\mathbf{g}_k||^2$. In particular, $m_Z = \bm{\mathsf{a}}_k^H\mathbf{R}_C\bm{\mathsf{a}}_k+ \mathsf{b}^2Tr(\mathbf{R}\mathbf{R}_C)$, and $v_Z=\mathsf{b}^4Tr((\mathbf{R}\mathbf{R}_C)^2) + 2\mathsf{b}^2\bm{\mathsf{a}}_k^H\mathbf{R}_C\mathbf{R}\mathbf{R}_C\bm{\mathsf{a}}_k$, where $\bm{\mathsf{a}}_k=\sqrt{\beta}\sqrt{\frac{\kappa}{1+\kappa}}\mathbf{h}_{k,LoS}$, $\mathsf{b}=\sqrt{\beta}\sqrt{\frac{1}{1+\kappa}}$, and $\mathbf{R}_C=\mathbf{C}\mathbf{C}^H$, where $\mathbf{C} = \mathbf{B}\mathbf{\Psi}_1$.}

\textit{Proof}: See Appendix A.

\textit{\textbf{Asymptotic at high SNR}:} At the high SNR regime (i.e., $\xi_k \rightarrow 0$), we can adopt an approximation of $\gamma(c,x) \overset{x \rightarrow 0}{\approx} x^c/c$ for asymptotic expression. In particular, we have 
\begin{align} \label{O4}
P_{out,k}^{\infty}=\frac{1}{\Gamma(\vartheta_k+1)}\left(\frac{\xi_k}{\varkappa_k}\right)^{\vartheta_k}. 
\end{align}

\noindent \textit{\textbf{Remark}}: The approximation of $Z=||\mathbf{g}_k||^2$ as gamma distribution via moment matching is analytically convenience. In fact, the OP expression in \eqref{O3} is simple for evaluation. Also, the diversity order is obtained as $d=-\lim_{\bar{\gamma} \to \infty} \frac{\log P_{out,k}(\bar{\gamma})}{\log \bar{\gamma}}=\vartheta_k$ (cf. \eqref{O4}). We also note that $Z = ||\mathbf{g}_k||^2 = \mathbf{h}_k^H\mathbf{R}_C\mathbf{h}_k$, which is non-central quadrature form. Thus, an approximation can be attained via saddle point approximation (SPA) technique \cite{Gur}. Specifically, by computing the cumulant generating function (CGF) of $Z$, and then applying the Lugannani–Rice formula \cite{LR}, we obtain the following result. 

\noindent\textbf{\textit{Theorem 2:}} \textit{The approximate expression of the OP for the $k$-th user by using a SPA technique is given as}
\begin{align} \label{O5}
P_{out,k}=\Phi(\omega_k)+\phi(\omega_k)\left(\frac{1}{\omega_k}-\frac{1}{\varpi_k}\right), 
\end{align}
\textit{where $\Phi(\cdot)$ and $\phi(\cdot)$ denote the CDF and PDF of standard normal distribution. Also, $\omega_k=sign(s^{\star})\sqrt{2(s^{\star}\xi_k-\mathcal{C}_g(s^{\star}))}$, $\varpi_k=s^{\star}\sqrt{\mathcal{C}_g^{''}(s^{\star})}$, where $sign(x)$ is 1 if $x\geq 0$ and 0 if $x<0$, $s^{\star}$ is the solution of the saddlepoint equation $\mathcal{C}_g^{'}(s)=\xi_k$, and $\mathcal{C}_g(s)$ is the CGF function shown in Appendix B.}

\textit{Proof}: See Appendix B.

\noindent Performance evaluation of these two approximation approaches are provided in Section V. 

\section{Energy Efficiency Maximization}
\subsection{Optimization Problem Formulation}
Energy efficiency (EE) is one of the key performance metrics in HAPS-based wireless systems. In this section, the EE is defined as the ratio of the sum rate across the users over the total power consumption in the system \cite{Per}. Also, equal power allocation among unicast and multicast streams is assumed for simplicity, i.e., $p_k = P/(K+1), k=0,1,...,K$. We focus on joint optimization of the total transmit power and the SIM phase-shifts for the maximal EE. To this end, the optimization problem is formulated as 
\begin{subequations}
\begin{alignat}{2}
(P0):\hspace{1pt} &\max_{P,\{\theta_n^l\}} &\qquad& EE(P,\mathbf{\Theta})= \frac{W_B\sum_{k=0}^K R_k(P,\mathbf{\Theta})}{P/\eta+P_c} \label{eq:optProb0}\\
&\text{subject to} &      & P \leq P_{max} \label{eq:constraint21}\\
&                  &      & R_k(P,\mathbf{\Theta}) \geq R_k^{th},\forall k \in \mathcal{K} \label{eq:constraint22} \\
&                  &      & R_0(P,\mathbf{\Theta}) \geq R_0^{th} \label{eq:constraint23} \\
&                  &      & \theta_n^l \in [0,2\pi), \forall n \in \mathcal{N}, \forall l \in \mathcal{L} \label{eq:constraint24} \\
&                  &      & P \geq 0, \label{eq:constraint25}
\end{alignat}
\end{subequations} 
\noindent where $W_B$ is the transmission bandwidth, $\eta \in (0,1)$ is the efficiency of power amplifiers (PA), and $P_c$ denotes the total circuit power consumption. This power can be computed as $P_c = N \times L \times P_{SIM}+ K\times P_{HAPS, RF}+ P_{HAPS,BB}+K\times P_{c,User}$, where $P_{SIM}$ is the power consumption of each SIM element, $P_{HAPS,RF}$ is the power consumed by each RF chain (excluding the PA) at HAPS, $P_{HAPS,BB}$ is the total power consumption of the baseband circuitry at HAPS\footnote{This power represents the aggregate power, including the power consumed by the SIM controller.}, and $P_{c, User}$ is the total power consumption at each ground user. Also, the constraints of \eqref{eq:constraint22} and \eqref{eq:constraint23} guarantee the quality of service (QoS) for each ground user. It is noted that the problem $(P0)$ is a non-convex optimization since the optimization variables of $P$ and $\mathbf{\Theta}$ are coupled within the non-convex objective function and the constraints of \eqref{eq:constraint22} and \eqref{eq:constraint23}. In what follows, we propose methods to solve this problem.

\subsection{Alternative Optimization (AO) Approach}
To tackle the above challenge, we propose an AO method to obtain a suboptimal solution of the problem $(P0)$ in this section. Specifically, the joint optimization problem is decomposed into two subproblems: optimization of total transmit power given SIM phase-shifts and optimization of phase-shifts given the transmit power. The AO algorithm will be completed when a convergence threshold or a maximum number of iterations is reached\footnote{Note that an iterative optimization approach and a penalty-based method used to handle coupled non-convex variables in a multi-RIS-based radar system was studied in \cite{Esm}.}.
  
\subsubsection{Optimize Total Transmit Power Given Phase-Shifts}
Given the SIM phase-shifts $\mathbf{\Theta}$, the optimization problem $(P0)$ is simplified to 
\begin{subequations}
\begin{alignat}{2}
(P1):\hspace{1pt} &\max_{P} &\qquad& EE(P) \label{eq:optProb1}\\
&\text{subject to} &      & \eqref{eq:constraint21}, \eqref{eq:constraint22}, \eqref{eq:constraint23}, \eqref{eq:constraint25}. \label{eq:constraint255}
\end{alignat}
\end{subequations} 

\noindent For notational convenience, let us denote $p=P/(K+1)$. We first simplify the constraints in \eqref{eq:constraint255}. The unicast rate constraint in \eqref{eq:constraint22} can be rewritten as (cf. \eqref{S10}-\eqref{S11})
\begin{align} \label{PA0}
P \geq (K+1)\max_{k}\frac{(2^{R_k^{th}}-1)N_0}{||\mathbf{g}_k||^2}\triangleq P^{unicast}_{min}.
\end{align}
\noindent Also, the multicast constraint \eqref{eq:constraint23} is rewritten as (cf. \eqref{S9}, \eqref{S12})
\begin{align} \label{PA1}
P \geq (K+1)\max_{k}\frac{(2^{R_0^{th}}-1)N_0}{||\mathbf{g}_k||^2(2-2^{R_0^{th}})}\triangleq P^{multicast}_{min},
\end{align}
\noindent where $R_0^{th} \leq 1 (bits/s/Hz)$ for feasible $P$. By denoting $P_{min} \triangleq \max\{P^{unicast}_{min}, P^{multicast}_{min}\}$, we can express all the constraints in \eqref{eq:constraint255} as $P_{min} \leq P \leq  P_{max}$. Thus, the problem ($P1$) is now equivalent to 
\begin{subequations}
\begin{alignat}{2}
(P1A):\hspace{1pt} &\max_{P} &\qquad& EE(P)= \frac{W_B\sum_{k=0}^K R_k(P)}{P/\eta+P_c}. \label{eq:optProb1A} \\
&\text{subject to} & & P_{min} \leq P \leq  P_{max}. \label{eq:constraint256}
\end{alignat}
\end{subequations} 
Next, let us consider the objective function $EE(P)$. We note that all the rate components $R_k(P)$ are concave with respect to (w.r.t.) $P$. Indeed, for the unicast term, we have $\frac{\partial^2 R_k(P)}{\partial P^2}=-\frac{(||\mathbf{g}_k||^2/N_0)^2}{(1+||\mathbf{g}_k||^2P/(K+1)N_0)^2\ln 2}<0$. Also, for the multicast term $R_0=\min_{k=1,2,...,K}\{R_0^k\}$, we have $\frac{\partial^2 R_0^k(P)}{\partial P^2}=-\frac{(||\mathbf{g}_k||^2/N_0)^2(3+4||\mathbf{g}_k||^2P/(K+1)N_0)}{(1+2||\mathbf{g}_k||^2P/(K+1)N_0)^2(1+||\mathbf{g}_k||^2P/(K+1)N_0)^2\ln 2}<0$. Since the minimum of concave functions, as well as the sum of concave functions, is concave, the nominator of $EE(P)$ is concave w.r.t. $P$. On the other hand, the denominator of $EE(P)$ is affine and strictly increasing w.r.t. $P$. Thus, the objective function $EE(P)$ is quasi-concave w.r.t. $P$. As a result, we can use efficient one-dimensional search methods, e.g., golden-section search \cite{Sca}, for finding out the optimal value $P^{\star}$ over the range $[P_{min},P_{max}]$. A summary of an algorithm for solving the problem $(P1)$ is provided in \textit{Algorithm 1}.

\begin{algorithm}[!t]
\caption{Algorithm for optimizing transmit power}
\begin{algorithmic}[1]
		\STATE \textbf{Input}: Channel coefficients $\{\textbf{h}_k^H\}$, phase-shifts $\mathbf{\Theta}$
    \STATE Calculate the minimum power constraint $P_{min}$
    \STATE Initialize: $p_l=P_{min}, p_u=P_{max}$, $\delta_p=p_u-p_l$; $\aleph=(\sqrt{5}-1)/2$, $i=1$, and number of iterations $I_{PA}$
    \STATE $\varsigma_1=p_l+\aleph\delta_p; \varsigma_2=p_u-\aleph\delta_p$ 
		\REPEAT 
        \IF {$EE(\varsigma_1) < EE(\varsigma_2) $}
        \STATE $p_u=\varsigma_1; \varsigma_1=\varsigma_2; \delta_p=\aleph\delta_p; \varsigma_2=p_u-\aleph\delta_p$
        \ELSE
        \STATE $p_l=\varsigma_2; \varsigma_2=\varsigma_1; \delta_p=\aleph\delta_p; \varsigma_1=p_l+\aleph\delta_p$
        \ENDIF
				\STATE Update $i=i+1$
    \UNTIL $(i>I_{PA})$ or the solution converges 
		\STATE \textbf{Output}: Optimal transmit power $P^{\star}=(p_u+p_l)/2$.
\end{algorithmic}
\end{algorithm}

\subsubsection{Optimize Phase-Shifts Given Transmit Power}
Given the transmit power value $P$, the optimization problem $(P0)$ is simplified to 
\begin{subequations}
\begin{alignat}{2}
(P2):\hspace{1pt} &\max_{\{\theta_n^l\}} &\qquad& EE(\mathbf{\Theta}) \label{eq:optProb2}\\
&\text{subject to} &      & \eqref{eq:constraint22}, \eqref{eq:constraint23}, \eqref{eq:constraint24} \nonumber. 
\end{alignat}
\end{subequations} 

\noindent Note that the denominator of $EE(\mathbf{\Theta})=\frac{W_B\sum_{k=0}^K R_k}{P/\eta+P_c}$ is a constant for given transmit power $P$. Thus, we only need to maximize the numerator of $EE(\mathbf{\Theta})$. In other words, $(P2)$ is equivalent to  
\begin{subequations}
\begin{alignat}{2}
(P2A):\hspace{1pt} &\max_{\{\theta_n^l\}} &\qquad& R_{sum}=\sum_{k=0}^K R_k \label{eq:optProb2A}\\
&\text{subject to} &      & \eqref{eq:constraint22}, \eqref{eq:constraint23}, \eqref{eq:constraint24} \nonumber. 
\end{alignat}
\end{subequations}

\noindent For phase-shift optimization in SIM-based systems, some existing works have used a computationally efficient gradient ascent method to update the phase-shifts, e.g., \cite{An}-\hspace{1pt}\cite{Pap1}. However, these studies only considered unconstrained problems. To take into account the rate constraints in \eqref{eq:constraint22} and \eqref{eq:constraint23}, we consider a penalty-based approach, i.e., penalty added to the objective when constraints are violated. Specifically, the optimization uses projected gradient ascent (PGA) for phase updates and adds a penalty to the sum-rate when constraints are violated. The penalized objective is formulated as
\begin{align} \label{GA0}
\tilde{R}_{sum}=&\sum_{k=0}^KR_k-\rho \sum_{k=1}^K\max(0,R_k^{th}-R_k) 
\end{align}
\noindent where $\rho$ a penalty coefficient. Thus, the optimization problem now becomes
\begin{subequations}
\begin{alignat}{2}
(P2B):\hspace{1pt} &\max_{\{\theta_n^l\}} &\qquad& \tilde{R}_{sum} \label{eq:optProb2B}\\
&\text{subject to} &      & \eqref{eq:constraint24} \nonumber. 
\end{alignat}
\end{subequations}

\noindent To proceed with the PGA, we calculate the partial derivative of $\tilde{R}_{sum}$ w.r.t. the phase-shift $\theta_n^l$, i.e., 
\begin{align} \label{GA1}
\frac{\partial \tilde{R}_{sum}(\theta_n^l)}{\partial \theta_n^l}=(1+\rho_0)\frac{\partial R_0(\theta_n^l)}{\partial \theta_n^l}+\sum_{k=1}^K(1+\rho_k)\frac{\partial R_k(\theta_n^l)}{\partial \theta_n^l},
\end{align}
\noindent where $\rho_k = \rho$ if $R_k<R_k^{th}$, and $\rho_k = 0$ if $R_k\geq R_k^{th}, \forall k = 0,1,...,K$. Given the partial derivatives of $\tilde{R}_{sum}(\theta_n^l)$, the phase-shifts can be updated as 
\begin{align} \label{GA2}
\theta_n^l =\theta_n^l + \nu \frac{\partial \tilde{R}_{sum}(\theta_n^l)}{\partial \theta_n^l}, \forall n \in \mathcal{N}, \forall l \in \mathcal{L},
\end{align}
\noindent where $\nu>0$ is the Armijo step size that can be obtained via the backtracking line search at each iteration \cite{An}-\hspace{1pt}\cite{An1}. 

Let us now calculate the derivatives of $R_0(\theta_n^l)$ and $R_k(\theta_n^l)$. For the multicast rate, we have
\begin{align} \label{GA3}
\frac{\partial R_0}{\partial \theta_n^l}=\frac{1}{\ln 2}\times \frac{1}{1+\gamma_{k'}^{\mathcal{M}}} \times \frac{\partial \gamma_{k'}^{\mathcal{M}}}{\partial \theta_n^l},
\end{align}
\noindent where $k' = \arg \min_{k=1,2,...,K}\{\log_2(1+\gamma_k^{\mathcal{M}})\}$, and (cf. \eqref{S9})
\begin{align} \label{GA4}
\frac{\partial \gamma_{k'}^{\mathcal{M}}}{\partial \theta_n^l}=\frac{p_0N_0}{(||\mathbf{g}_{k'}||^2p_{k'}+N_0)^2}\times \frac{\partial}{\partial \theta_n^l}||\mathbf{g}_{k'}||^2.
\end{align}

\noindent Similarly, for the unicast rates, we have (cf. \eqref{S10})
\begin{align} \label{GA5}
\frac{\partial R_k}{\partial \theta_n^l}=\frac{1}{\ln 2}\times \frac{1}{1+\gamma_k^{\mathcal{U}}}\times \frac{p_k}{N_0}\times \frac{\partial}{\partial \theta_n^l}||\mathbf{g}_k||^2.
\end{align}

\noindent To evaluate \eqref{GA4} and \eqref{GA5}, we need to derive the expression of $\frac{\partial ||\mathbf{g}_k||^2}{\partial \theta_n^l}, \forall k \in \mathcal{K}$. Specifically, we can express 
\begin{align} \label{GA6}
\frac{\partial ||\mathbf{g}_k||^2}{\partial \theta_n^l}=2\mathbb{R}\Big\{\frac{\partial \mathbf{g}_k}{\partial \theta_n^l}\mathbf{g}_k^H\Big\},
\end{align}

\noindent where $\mathbb{R}\{z\}$ is the real part of $z$. Let us rewrite $\mathbf{B}$ as $\mathbf{B}=\mathbf{B}_{l,1}\mathbf{\Theta}_l\mathbf{B}_{l,2}$, where $\mathbf{B}_{l,1} \triangleq \mathbf{\Theta}_L\mathbf{\Psi}_L\mathbf{\Theta}_{L-1}\mathbf{\Psi}_{L-1} \cdots \mathbf{\Theta}_{l+1}\mathbf{\Psi}_{l+1}$ and $\mathbf{B}_{l,2} \triangleq \mathbf{\Psi}_l\mathbf{\Theta}_{l-1}\mathbf{\Psi}_{l-1} \cdots \mathbf{\Psi}_2\mathbf{\Theta}_1$ (cf. \eqref{S2}). Then, we have
\begin{align} \label{GA7}
\frac{\partial \mathbf{g}_k}{\partial \theta_n^l}=\mathbf{h}_k^H\mathbf{B}_{l,1}\frac{\partial \mathbf{\Theta}_l}{\partial \theta_n^l}\mathbf{B}_{l,2}\mathbf{\Theta}_1, 
\end{align}  
\noindent where $\frac{\partial \mathbf{\Theta}_l}{\partial \theta_n^l}=je^{j\theta_n^l}\mathbf{E}_n$, and $\mathbf{E}_n =\operatorname{diag}(\mathbf{e}_n)$, where $\mathbf{e}_n = [0,0,\cdots, 0,1,0,\cdots,0]$ is the vector with a 1 at the $n^{th}$ position and 0 elsewhere. As a result, we obtain 
\begin{align} \label{GA8}
\frac{\partial ||\mathbf{g}_k||^2}{\partial \theta_n^l}=2\mathbb{R}\{je^{j\theta_n^l}\mathbf{h}_k^H\mathbf{B}_{l,1}\mathbf{E}_n\mathbf{B}_{l,2}\mathbf{\Psi}_1\mathbf{\Psi}_1^H\mathbf{B}^H\mathbf{h}_k\}.
\end{align}
\noindent By substituting the result in \eqref{GA8} into \eqref{GA4} and \eqref{GA5}, we obtain the expression of $\frac{\partial \tilde{R}_{sum}(\theta_n^l)}{\partial \theta_n^l}$ as shown in \eqref{GA9} (see on the top of the next page). A summary of the PGA-based algorithm for optimizing the phase-shifts is provided in \textit{Algorithm 2}. 

\begin{figure*}
\begin{align} \label{GA9}
\frac{\partial \tilde{R}_{sum}(\theta_n^l)}{\partial \theta_n^l}=&(1+\rho_0)\times \frac{2}{\ln 2}\times \frac{1}{1+\gamma_{k'}^{\mathcal{M}}}\times \frac{p_0N_0}{(||\mathbf{g}_{k'}||^2p_{k'}+N_0)^2}\times \mathbb{R}\{je^{j\theta_n^l}\mathbf{h}_{k'}^H\mathbf{B}_{l,1}\mathbf{E}_n\mathbf{B}_{l,2}\mathbf{\Psi}_1\mathbf{\Psi}_1^H\mathbf{B}^H\mathbf{h}_{k'}\} \nonumber\\
&+\sum_{k=1}^K(1+\rho_k)\times \frac{2}{\ln 2} \times \frac{1}{1+\gamma_k^{\mathcal{U}}}\times \frac{p_k}{N_0}\times \mathbb{R}\{je^{j\theta_n^l}\mathbf{h}_k^H\mathbf{B}_{l,1}\mathbf{E}_n\mathbf{B}_{l,2}\mathbf{\Psi}_1\mathbf{\Psi}_1^H\mathbf{B}^H\mathbf{h}_k\}
\end{align} 
\hrulefill
\vspace*{-8pt}
\end{figure*}

\begin{algorithm}[!t]
\caption{Algorithm for optimizing phase-shifts}
\begin{algorithmic}[1]
		\STATE \textbf{Input}: Channel coefficients $\{\textbf{h}_k^H\}$, transmit power $P$
    \STATE \textbf{Initialize}: Phase-shifts $\{\theta_n^l\}$, $i=1$, a maximum number of iterations $I_{PS}$, tolerance $\varepsilon >0$
		\REPEAT 
        \STATE Compute the gradient using \eqref{GA9} 
				\STATE Update $\theta_n^l$ using \eqref{GA2}
				\STATE Project back to the feasible range, i.e., $\theta_n^l \leftarrow \mod(\theta_n^l,2\pi)$ 
			  \STATE Update $i = i+1$
    \UNTIL $(i>I_{PS})$ or the solution converges (i.e., $|\tilde{R}_{sum}(i+1)-\tilde{R}_{sum}(i)|<\varepsilon$)
		\STATE \textbf{Output}: Optimal phase-shifts $\{\mathbf{\Theta}^{\star}\}$.
\end{algorithmic}
\end{algorithm}

\subsubsection{Joint Optimization of Transmit Power and Phase-Shifts}
To realize the joint optimization of transmit power and phase-shifts for maximal EE, we alternatively perform \textit{Algorithm 1} and \textit{Algorithm 2} until convergence. A complete AO-based algorithm for solving the problem $(P0)$ is provided in \textit{Algorithm 3}. Regarding computational complexity of the proposed AO-based algorithm, let us evaluate the complexity of each subproblems. For the transmit power optimization problem, its complexity depends on the complexity the one-dimensional search method. When golden-section search  is used, it has a complexity of $\mathcal{O}(K\log(\Delta_P/\epsilon))$, where $\Delta_P = P_{max}-P_{min}$, and $\epsilon$ denotes target absolute accuracy. Also, for the PGA algorithm, the complexity for evaluating the gradient in \eqref{GA1} for all $N\times L$ phase-shifts $\theta_n^l$ is $\mathcal{O}(I_{PS}LKN^2)$, where $N>>K$. As a result, the total complexity of the AO-based algorithm is $\mathcal{O}(I_{AO}[K\log(\Delta_P/\epsilon)+I_{PS}LKN^2])$, where $I_{AO}$ is the number of iterations of the main loop\footnote{The computational complexity of AO-based joint optimization of power allocation and phase-shift optimization in a multi-user SIM-aided MISO system is $\mathcal{O}(I_{AO}K^2(4N+3)[I_{IWF}+2NLI_{PS}])$ \cite{An}, whereas that of cell-free massive MIMO with SIM is $\mathcal{O}(I_{AO}[3K^2N^2L^2+K^2LN^2])$ \cite{Hu}.}. This indicates a computational feasibility and reveals the impacts of the key system parameters on the overall complexity.

\begin{algorithm}[!t]
\caption{AO-based algorithm for joint optimization}
\begin{algorithmic}[1]
		\STATE \textbf{Input}: Channel coefficients $\{\textbf{h}_k^H\}$
    \STATE \textbf{Initialize}: Phase-shifts $\mathbf{\Theta}^{(0)}$, $i=1$, a maximum number of iterations $I_{AO}$, tolerance $\varepsilon >0$, 
		\REPEAT 
				\STATE Use \textit{Algorithm 1} to find the optimal transmit power $P^{(i)}$ given phase-shifts $\mathbf{\Theta}^{(i-1)}$ 
				\STATE Use \textit{Algorithm 2} to find the optimal phase-shifts $\mathbf{\Theta}^{(i)}$ given the power $P^{(i)}$
				\STATE Update $i = i+1$
    \UNTIL $(i>I_{AO})$ or the solution converges (i.e., $|EE(i+1)-EE(i)|<\varepsilon$)
		\STATE \textbf{Output}: Optimal transmit power $\{P^{\star}\}$ and phase-shifts $\{\theta_n^{l,\star}\}$.
\end{algorithmic}
\end{algorithm}

\subsection{Unsupervised Deep Neural Networks (DNN) Approach}
In this section, we exploit recent advances in machine learning for solving optimization problems. In general, supervised learning often relies on time-consuming optimization techniques for a generation of training data set since labeling is required. In \cite{Gao}, an unsupervised DNN method was developed to solve a rate maximization problem in a single-user RIS-based system. It maintains most of the performance while significantly reducing computational complexity compared to a conventional optimization based approach. Motivated by this, we propose an unsupervised learning approach to solve the EE maximization problem (P0)\footnote{An investigation of other learning-based methods (i.e., DRL) is an interesting direction and is left for future work.}. 

\begin{figure}[t] 
	\centering{\includegraphics[width=0.45\textwidth]{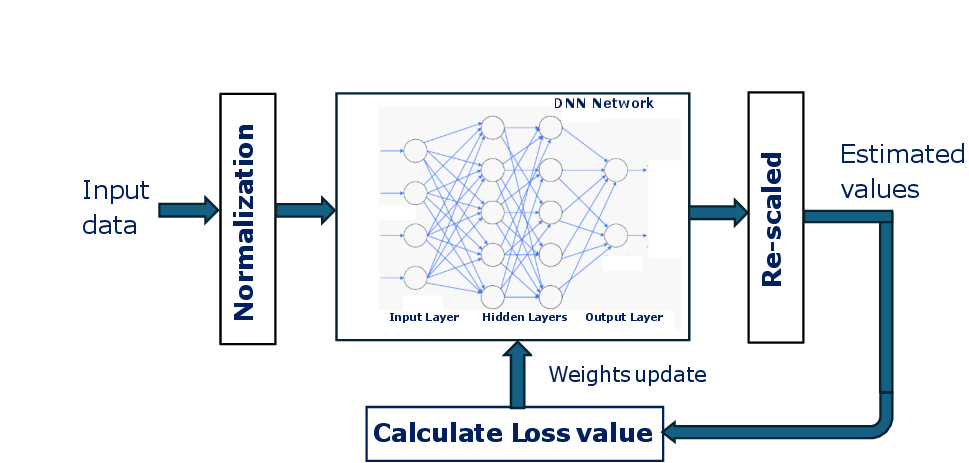}}
	\caption{An unsupervised DNN network architecture.}
	\label{Fig2}
\end{figure}

\subsubsection{DNN Architecture}
We define a fully-connected (FC) feed-forward DNN as shown in Fig. 2. It consists of one input layer, five hidden layers, and one output layer. Batch normalization layers are used to prevent overfitting and promote efficient training. Also, rectified linear unit (\textit{ReLU}) and sigmoid layer (\textit{sigmoidLayer}) activations are used in the hidden layers and the output layer, respectively. The input to the DNN is the channel coefficients from the SIM to $K$ ground users, i.e., $\mathbf{H}=[\mathbf{h}_1, \mathbf{h}_2,\cdots, \mathbf{h}_K]$. Note that the real and imaginary parts of $\mathbf{H}$ need to be separated as two features since the DNN is designed to process real numbers. Also, the output is the transmit power and the phase shifts of the SIM. As a result, the input and output dimensions of the DNN are $2NK$ and $NL+1$, respectively. It is worth noting that the complexity of the DNN can be evaluated in terms of floating point operations (FLOPS). Specifically, for a fully-connected (FC) layer with $N_{i}$ inputs and $N_{o}$ outputs, the number of FLOPS is approximately $2N_{i}N_{o}$ \cite{Mol}. Thus, the number of FLOPS of all FC layers in the DNN is $8J^2+4NKJ+2J(NL+1)$, where $J$ denotes the number of neurons in each hidden layer. Taking the batch normalization, ReLU activation, and sigmoid activation into account, the total number of FLOPS is $8J^2+4NKJ+2J(NL+1)+25J+4NL+4$. For large $J$, the complexity is dominated by $\mathcal{O}(8J^2+4NKJ+2NLJ)$.   

\subsubsection{Loss Function}
In conventional supervised learning for regression, a loss function is commonly defined as a mean square error (MSE) between the true values and the model's predictions. A low loss value will indicate that the model produces accurate predictions. In contrast, a loss function in unsupervised learning is set to be the negative of the objective function of the optimization problem. This is because the DNN is trained in the direction of minimizing the loss function, which corresponds exactly to an increasing of the objective function. Specifically, in our system, after obtaining the transmit power and phase shifts of each elements in the SIM, the EE can be computed based on \eqref{eq:optProb0}. The loss function used to update the parameters of the DNN network is defined as
\begin{align} \label{L0}
Loss=-\frac{1}{S}\sum_{s=1}^S \Big[EE(P_s,\mathbf{\Theta}_s,\mathbf{H}_s)-\rho u(\varrho_s)\Big], 
\end{align}
\noindent where $S$ is the number of training data samples in a minibatch, $\rho$ is a parameter that reflects the penalty level applied when the constraints are violated, $u(x) = ReLU(x)$, and $\varrho_s=\max\left(0,P_s - P_{max}\right)+\max(0,R_0^{th}-R_0^s)+\sum_{k=1}^K\max(0,R_k^{th}-R_k^s)$ is the penalty indicator. Note that if any of the constraints is not satisfied, the loss value is large due to the addition of positive penalty term. On the other hand, when all the constraints are satisfied (i.e., $\varrho_s=0$), the penalty term equals to zero, and thus having no impacts on the loss value. Furthermore, the parameter $\rho$ should be chosen large enough to guarantee strict satisfaction of the constraints\footnote{The penalty parameter is selected using a heuristic based on constraint violation levels.}.

\subsubsection{Unsupervised Learning-based Training and Testing}
In the training phase, channel coefficients are estimated to construct samples for training. Recall that the unsupervised learning approach does not requires labeling. Thus, the cost for a generation of a data set is low. To improve stability and performance, normalization is implemented before feeding data to the DNN network. Also, adaptive moment estimation (Adam) optimizer is utilized to update network parameters. By using the loss function defined above to update the model’s weights, the DNN network then learns how to optimize the EE. Once the DNN is trained offline, it then is deployed online for real-time prediction. Note that during the testing the DNN is evaluated on channel samples that are disjoint from the training set.

\section{Simulation Results and Discussions}
In this section, we provide simulation results to validate the OP analysis and demonstrate the efficacy of the proposed approaches for the EE maximization problem. Unless otherwise specified, the simulation parameters are listed in Table I. The coordinate of the HAPS is $(0,0,21) (km)$, whilst those of the ground users are $(2,-3,0), (5,-10,0), (12,1,0)$, and $(4,25,0) (km)$. For the unsupervised DNN approach, the number of neurons in each hidden layer is set to 256. Also, the batch size is set to 50, and the maximum number of training epochs is 2000.

\begin{table}[!t]
\scriptsize
\begin{center}
\captionsetup{font=scriptsize}
\caption{Simulation parameters.}
\vspace{10pt}
\begin{tabular}{|@{\hskip3pt}l@{\hskip3pt}|@{\hskip3pt}l@{\hskip3pt}|}
\hline
\hfil \textbf{Parameters} & \hfil \textbf{Values}\\
\hline
Operating frequency  & $f = 2.1$ (GHz) \\
Transmission bandwidth  & $W_B=20$ (MHz) \\
Altitude of HAPS & $h_{HAPS} = 21$ (km) \\
Efficiency of power amplifiers & $\eta=0.85$ \\
Circuit power consumption  & $P_{SIM}=10$, $P_{HAPS,BB}=40$ (dBm) \cite{Per}\\
& $P_{c, User}=10$, $P_{HAPS,RF}=30$ (dBm) \cite{Per} \\
Antenna gains & $G_t=5$, $G_r=0$ (dBi) \cite{An}-\hspace{1pt}\cite{An1}\\
Number of elements per layer & $N = 25$ \\ 
Number of SIM layers & $L = 3$ \\ 
Spacing between SIM units & $d_e = \lambda/2$ \cite{An}-\hspace{1pt}\cite{An1}\\ 
Size of each SIM element & $d_x = d_y = \lambda/2$ \cite{An}-\hspace{1pt}\cite{An1}\\ 
Thickness of the SIM & $T_{SIM} = 5\lambda$ \cite{An}-\hspace{1pt}\cite{An1}\\ 
Spacing between SIM layers & $d_{SIM}=T_{SIM}/L$ \cite{An}-\hspace{1pt}\cite{An1}\\ 
Rician factor & $\kappa = 4 $ \\ 
Number of users & $K = 4 $ \\ 
Noise power spectral density & $PSD = -174$ (dBm/Hz) \\
\hline
\end{tabular}
\end{center}
\end{table}

\begin{figure}[t] 
	\centering{\includegraphics[width=0.45\textwidth]{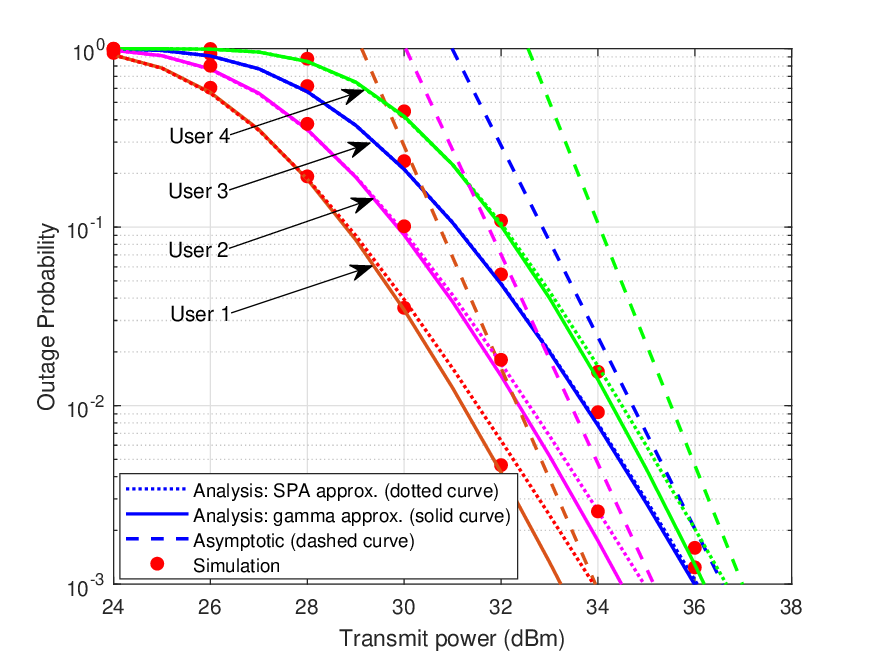}}
	\caption{OP versus transmit power ($\kappa =6$ dB).}
	\label{Fig3}
\end{figure}

\begin{figure}[t] 
	\centering{\includegraphics[width=0.45\textwidth]{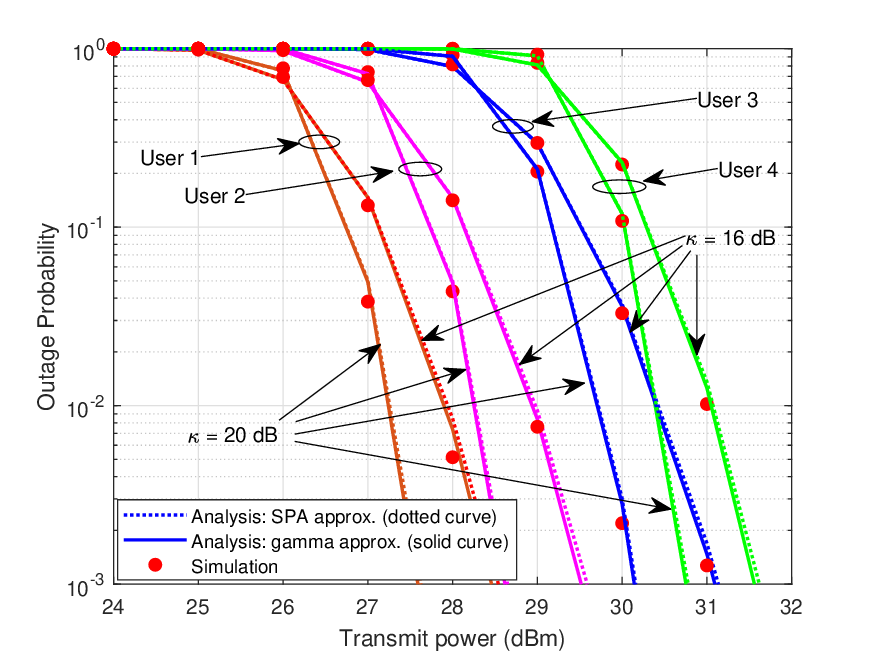}}
	\caption{OP versus transmit power ($\kappa =16, 20$ dB).}
	\label{Fig4}
\end{figure}
	
\subsection{Outage Probability}
The OP performance of each user versus the transmit power is shown in Fig. 3. Given that the OP expressions are derived for arbitrary values of power allocation and SIM phase-shifts, random values of phase-shifts and equal power allocation are considered here for demonstration. Additionally, the threshold rates for both the multicast and the unicast streams are set to $R_{th}=0.1$ (b/s/Hz). Note that these values are the same for both simulation and analysis curves for comparison. As expected, the OPs of all users become lower when the transmit power is higher. Moreover, it can be seen that the analysis curves match well with the simulation curves, which demonstrates the accuracy of the derived approximate expressions in \textit{Theorem 1} and \textit{Theorem 2}. An accuracy of the two approximation approaches is further illustrated in Fig. 4 where different values of the Rician factor, i.e., $\kappa = 16$ (dB) and $\kappa = 20$ (dB), are considered.

\begin{figure}[t] 
	\centering{\includegraphics[width=0.45\textwidth]{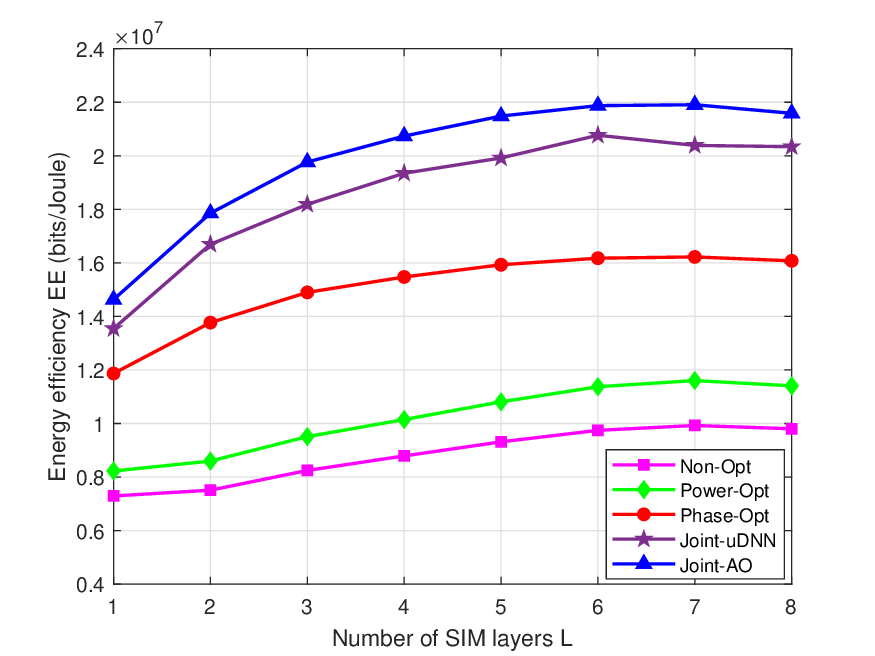}}
	\caption{EE versus a number of SIM layers ($P_{max}=20 W$).}
	\label{Fig5}
\end{figure}

\begin{figure}[t] 
	\centering{\includegraphics[width=0.45\textwidth]{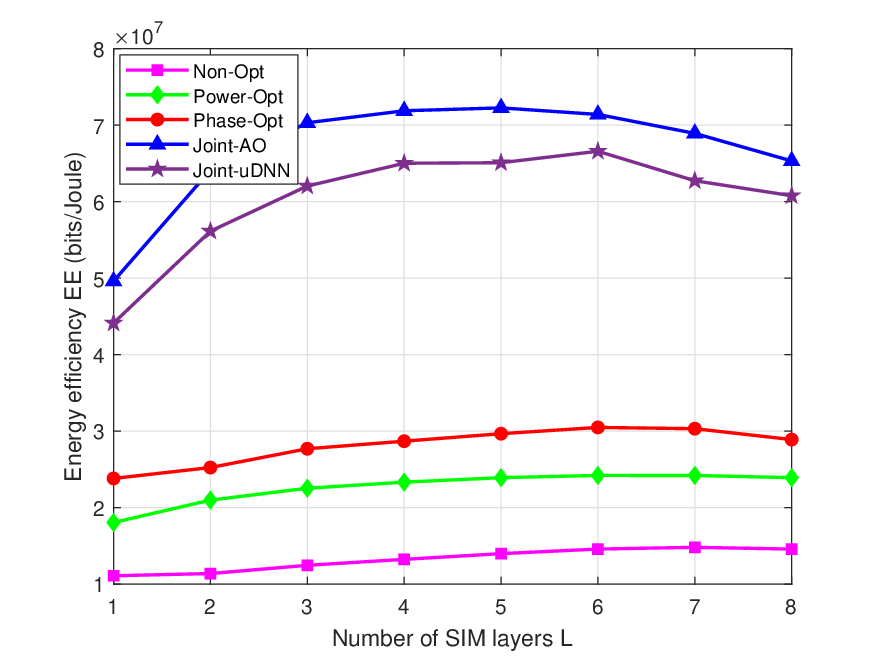}}
	\caption{EE versus a number of SIM layers ($P_{max}=20 W$, $P_{BB}=200 mW$, $P_{RF}=300 mW$).}
	\label{Fig6}
\end{figure}

\begin{figure}[t] 
	\centering{\includegraphics[width=0.45\textwidth]{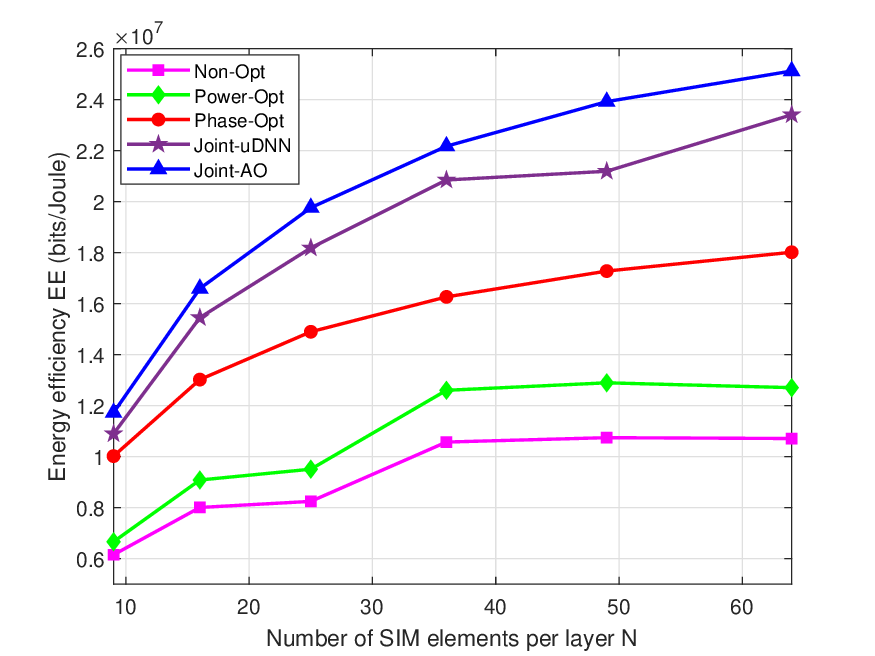}}
	\caption{EE versus a number of SIM elements per layer ($P_{max}=20 W$).}
	\label{Fig7}
\end{figure}

\begin{figure}[t] 
	\centering{\includegraphics[width=0.45\textwidth]{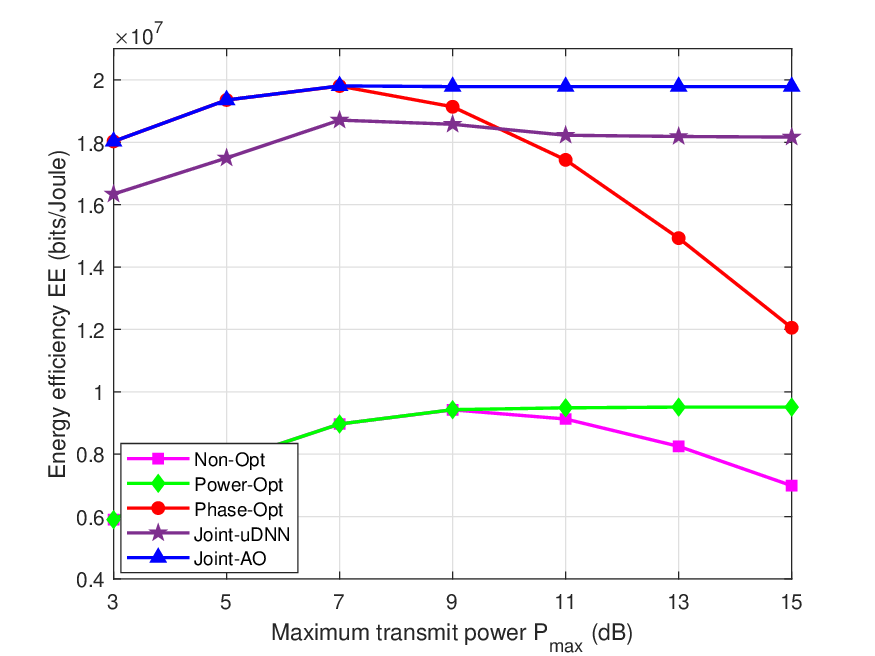}}
	\caption{EE versus maximal transmit power $P_{max}$.}
	\label{Fig8}
\end{figure}

\begin{figure}[t] 
	\centering{\includegraphics[width=0.45\textwidth]{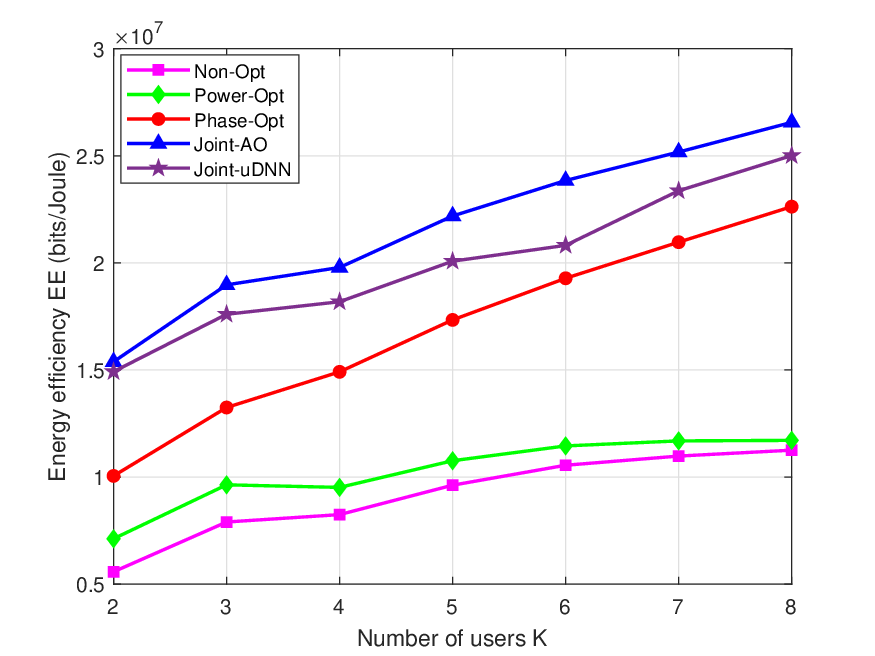}}
	\caption{Fig. 9. EE versus a number of ground users $K$.}
	\label{Fig9}
\end{figure}

\begin{figure}[t] 
	\centering{\includegraphics[width=0.45\textwidth]{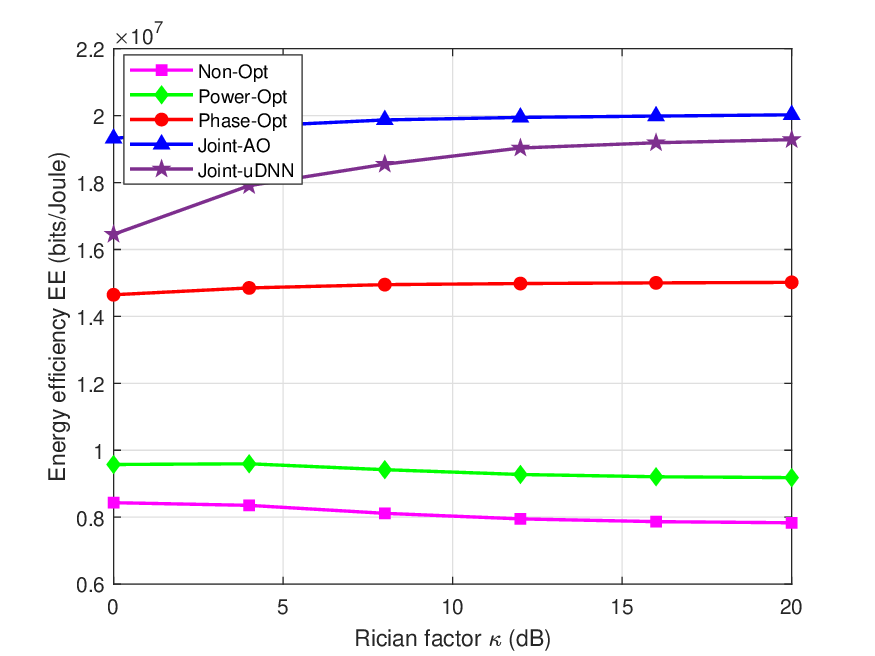}}
	\caption{Fig. 10. EE versus the Rician factor $\kappa$.}
	\label{Fig10}
\end{figure}

\begin{figure}[t] 
	\centering{\includegraphics[width=0.45\textwidth]{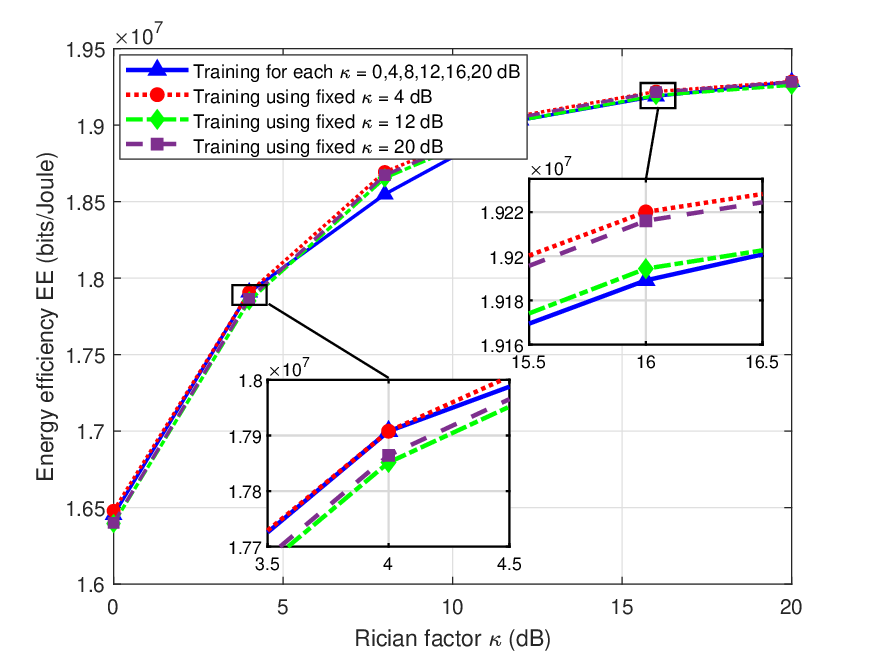}}
	\caption{EE versus $\kappa$ under different training approaches.}
	\label{Fig11}
\end{figure}

\begin{figure}[t] 
	\centering{\includegraphics[width=0.45\textwidth]{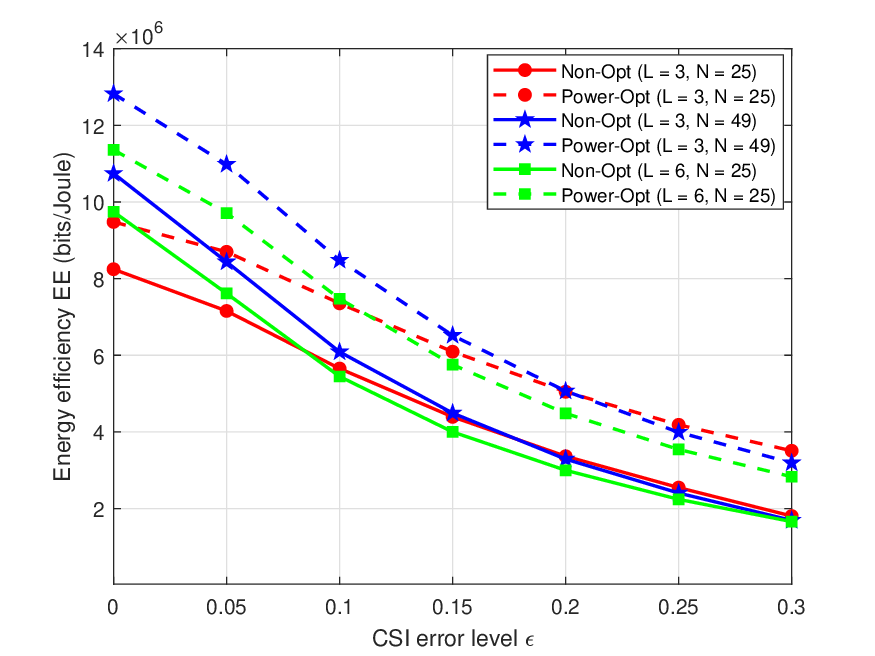}}
	\caption{EE versus CSI errors.}
	\label{Fig12}
\end{figure}

\subsection{Energy Efficiency}
We now evaluate the EE achieved with the proposed methods\footnote{Power values listed in Table I are converted to linear scale (i.e., Watts) prior to computing the energy efficiency using the standard conversion $P_{(W)} = 10^{(P_{(dBm)}-30)/10}$.}. Five schemes are considered, including: \textit{1)} \textit{Joint-AO}: Joint optimization of transmit power and phase shifts using alternative optimization; \textit{2)} \textit{Joint-uDNN}: joint optimization using unsupervised learning; \textit{3)} \textit{Power-Opt}: optimal transmit power with fixed phase-shifts; \textit{4)} \textit{Phase-Opt}: optimal phase-shifts with fixed transmit power; and \textit{5)} \textit{Non-Opt}: fixed transmit power with fixed phase-shifts. In these schemes, $P=P_{max}$ for fixed transmit power, and $\theta_n^l=\pi, \forall n \in \mathcal{N}, l \in \mathcal{L}$, for fixed phase-shifts. 

In Fig.5, we plot the EE performance versus a number of SIM layers $L$. First, it can be seen that when $L$ is increased, the EE first increases and then decreases. This is because, initially, the benefit in terms of capacity improvement thanks to additional SIM layers outweigh the impact of increased power consumption, resulting to improved EE. However, when $L$ becomes sufficiently large, the additional power consumption has a greater effect on the EE than the capacity improvement, leading to a decrease of the EE. Second, the joint optimization methods, i.e., \textit{Joint-AO} and \textit{Joint-uDNN}, can achieve higher EE than their counterparts. Obviously, the performance degradation highlight the necessity of the joint optimization of the transmit power and phase-shifts. Third, \textit{Joint-uDNN} attains slightly lower EE comparable to that of \textit{Joint-AO} while requiring significantly less running time, as demonstrated later in Table II. Fourth, it is worth noting that the stacked architecture (i.e., $L \geq 2$) outperforms the single-layer for a wide range of $L$. Due to constraints in terms of weight and size on HAPS platform, the number of SIM layer should not be too large. As a result, the deployment of SIM offers more benefits compared to the single-layer model. Similar observations can be made from Fig. 6, where an alternative set of power parameters that is consistent with HAPS platforms reported in \cite{Ji5} is adopted. Additionally, the EE performance versus the number of SIM elements per layer is shown in Fig. 7. These results highlight the necessity of carefully selecting the SIM parameters to achieve high EE performance.

Impacts of the maximum transmit power on the EE performance are shown in Fig. 8. The results show that when $P_{max}$ increases, the EE first increases. This can be explained by the fact that increasing transmit power initially boots the capacity, which enhances the EE. However, when the transmit power is sufficiently high, the capacity improvement becomes marginal, while the power consumption is large, leading to a reduction in the EE in the schemes with fixed transmit power. For the schemes with the optimized transmit power, transmit power is selected such that the EE value remains unchanged regardless of an increase of $P_{max}$. 

To further investigate impacts of system parameters on the EE performance, we plot in Fig. 9 and Fig. 10 the EE versus the number of ground users and the Rician factor $\kappa$, respectively. It can be seen from Fig. 9 that the EE increases when the number of users increases. This implies that the hybrid digital precoding and SIM-based beamforming can effectively mitigate the multi-user interference, leading to higher EE. Note that this behavior of the EE is similar to that in \cite{Per}, where a SIM-based MIMO system is investigated in terrestrial wireless environments. On the other perspective, Fig. 10 reveals that variations in the Rician factor $\kappa$ do not have a significant impact on the EE performance. Thanks to this behavior, the unsupervised DNN method trained using a specific value $\kappa$ can still be used for prediction under different channel conditions. This is illustrated in Fig. 11, where the EE performance achieved using a DNN trained at a particular $\kappa$ is close to that obtained when the training and testing channels share the same $\kappa$. These results imply that the proposed unsupervised DNN approach generalizes well across channel conditions with respect to the Rician factor.

In Fig. 12, we consider the EE in the presence of the CSI errors. In this case, the channel from the last layer of SIM to the $k$-th user is modeled as $\mathbf{h}_k=\sqrt{1-\epsilon^2}\mathbf{\tilde{h}}_k+\epsilon\mathbf{e}_k$, where $\epsilon$ denotes the accuracy of the CSI estimation (i.e., $\epsilon=0$ indicates no CSI errors), and $\mathbf{e}_k \sim \mathcal{CN}(0,\sigma_{e,k}^2)$ is the channel estimation error \cite{Ahn}. An extension of the system model considering imperfect CSI is provided in Appendix C. It can be seen that the EE performance degrades with an increase of CSI errors. This is because the presence of CSI errors reduces the SINR of both the multicast and unicast signals as well as the efficacy of beamforming at HAPS. As a consequence, the sum-rate is lower, and thus the achieved EE is lower (cf. (17.a)).

\begin{figure}[t] 
	\centering{\includegraphics[width=0.45\textwidth]{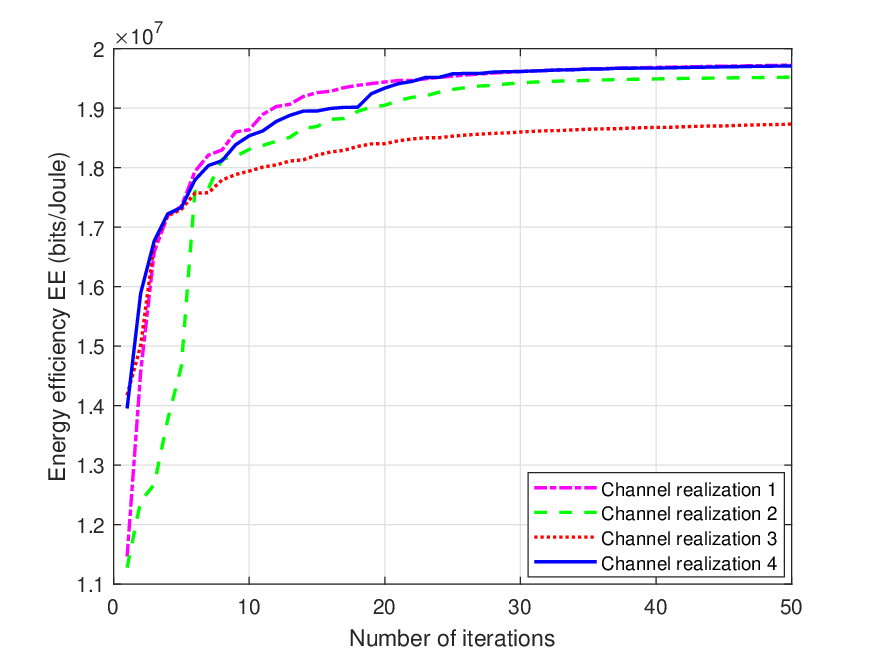}}
	\caption{EE versus a number of iterations.}
	\label{Fig13}
\end{figure}

\begin{figure}[t] 
	\centering{\includegraphics[width=0.45\textwidth]{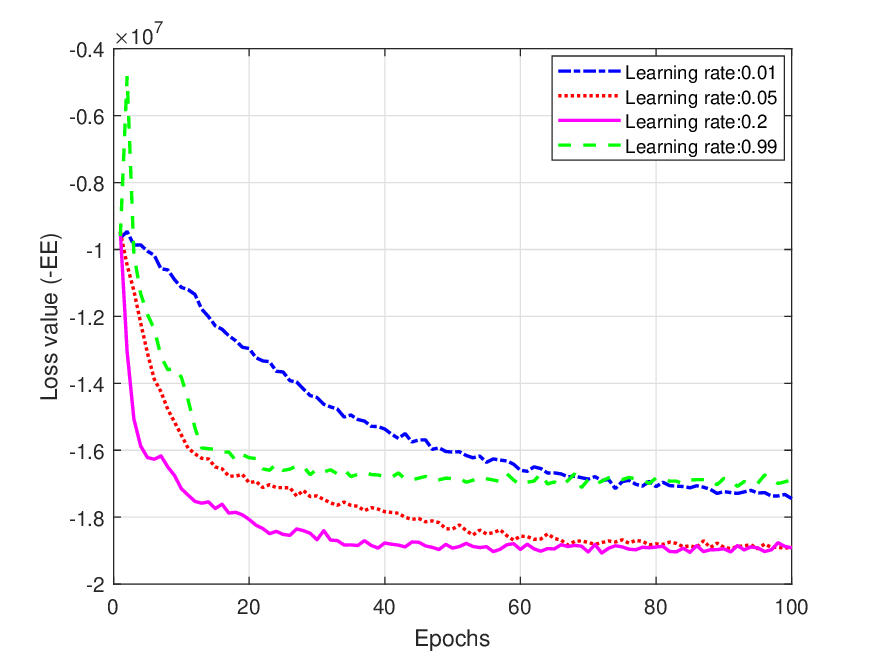}}
	\caption{Training loss versus a number of epochs.}
	\label{Fig14}
\end{figure}

\subsection{Analysis of Convergence and Complexity}
To demonstrate the convergence of the proposed AO algorithm, we plot in Fig. 13 the EE versus the number of iterations of four independent channel realizations. It can be seen that the AO algorithm converges quickly in all cases. Specifically, the maximum values of the EE can be attained after around 30 iterations.  

For the proposed unsupervised learning framework, we plot in Fig. 14 the evolution of the loss value over a number of epochs. As expected, the loss value decreases with an increase of training epochs. Also, a rate of the decrease depends the learning rate values. In particular, a learning rate of $l_r=0.2$ can achieve a good loss value compared to its counterparts, i.e., $l_r=0.01, 0.05, 0.99$. Additionally, the loss values are soon to reach the plateau, e.g., after about 35 epochs for the case of $l_r=0.2$. It is worth noting that, with the selected learning rate, the learned solutions closely match the AO results across different scenarios, such as channel conditions and system parameters, indicating the robustness and consistent convergence. 

Finally, in Table II, we compare the computational complexity in terms of the average running time of the two joint optimization methods. The proposed system under different parameters are considered. Simulations were performed on a laptop equipped with Intel Core i7-1185G7 CPU @3.0 GHz, 16 GB RAM, and running Windows 10 Education. Note that the time consumed for training in the \textit{Joint-uDNN} method is not considered here given that the training process is done offline. The results shown that \textit{Joint-uDNN} consumes significantly less time compared to \textit{Joint-AO}. As an example, when $L=3$ and $N=49$, \textit{Joint-uDNN} runs hundreds of times faster than \textit{Joint-AO}. This demonstrates the effectiveness of the proposed unsupervised learning method.

\begin{table}[!t]
\scriptsize
\begin{center}
\captionsetup{font=scriptsize}
\caption{Average Running Time Comparison (\textit{in seconds}).}
\vspace{10pt}
\begin{tabular}{|@{\hskip3pt}l@{\hskip3pt}|@{\hskip3pt}l@{\hskip3pt}|@{\hskip3pt}l@{\hskip3pt}|}
\hline
\hfil \textbf{SIM parameters} & \hfil \textbf{\textit{Joint-AO}} & \hfil \textbf{\textit{Joint-uDNN}}\\
\hline
$L=3, N=9$  & 0.0754 & 0.0014 \\
\hline
$L=2, N=25$  & 0.0913 & 0.0015 \\
\hline
$L=5, N=25$  & 0.2008 & 0.0015 \\
\hline
$L=3, N=49$  & 0.4182 & 0.0017 \\
\hline
\end{tabular}
\end{center}
\end{table}

\section{Conclusions}
We have proposed SIM-aided HAPS-based wireless systems for simultaneous multicast and unicast downlink transmissions in this work. The outage probability has been analytically characterized using gamma approximation and saddlepoint approximation methods. Also, the HAPS transmit power and the SIM phase-shifts have been optimized to achieve the maximal energy efficiency. The joint optimization problem is accomplished using the AO algorithm and unsupervised learning approaches. Furthermore, our results reveal the impacts of the number SIM elements per layer, the number of SIM layers, and the transmit power on system performance. As future research directions, it is worth investigating SIM-aided HAPS employing rate-splitting multiple access (RSMA), as well as scenarios with randomly distributed ground users.

\appendices
\section{Outage Probability via Gamma Approximation}\label{Appendix_A}
In this appendix, we will approximate $||\mathbf{g}_k||^2$ as a gamma distribution via moment matching. For notational convenience, let us define $\bm{\mathsf{a}}_k=\sqrt{\beta}\sqrt{\frac{\kappa}{1+\kappa}}\mathbf{h}_{k,LoS}$ and $\mathsf{b}=\sqrt{\beta}\sqrt{\frac{1}{1+\kappa}}$. Then, the channel coefficient $\mathbf{h}_k$ defined in \eqref{C1} can be rewritten as $\mathbf{h}_k=\bm{\mathsf{a}}_k+\mathsf{b} \mathbf{h}_{k,NLoS}$. Also, let us define $\mathbf{C} = \mathbf{B}\mathbf{\Psi}_1$ and $\mathbf{R}_C=\mathbf{C}\mathbf{C}^H$. We can express
\begin{align} \label{AA1}
Z = &||\mathbf{g}_k||^2 =||\mathbf{h}_k^H\mathbf{B}\mathbf{\Psi}_1||^2 = \mathbf{h}_k^H\mathbf{R}_C\mathbf{h}_k\nonumber\\
=& (\bm{\mathsf{a}}_k^H+\mathsf{b} \mathbf{h}_{k,NLoS}^H) \mathbf{R}_C(\bm{\mathsf{a}}_k+\mathsf{b} \mathbf{h}_{k,NLoS}) \nonumber\\
=& \bm{\mathsf{a}}_k^H\mathbf{R}_C\bm{\mathsf{a}}_k+\mathsf{b}^2\mathbf{h}_{k,NLoS}^H\mathbf{R}_C\mathbf{h}_{k,NLoS}+\mathsf{b} \mathbf{h}_{k,NLoS}^H\mathbf{R}_C\bm{\mathsf{a}}_k \nonumber\\
& +\mathsf{b}\bm{\mathsf{a}}_k^H \mathbf{R}_C\mathbf{h}_{k,NLoS}.
\end{align}
\noindent The first moment of $Z$ can be evaluated as
\begin{align} \label{AA2}
m_Z = \mathbb{E}\{Z\} = \bm{\mathsf{a}}_k^H\mathbf{R}_C\bm{\mathsf{a}}_k+ \mathsf{b}^2Tr(\mathbf{R}\mathbf{R}_C),
\end{align}
\noindent where $Tr(\cdot)$ denotes a trace of a matrix. Note that \eqref{AA2} is obtained based on the fact that $\bm{\mathsf{a}}_k$ and $\mathbf{R}_C$ are deterministic values, and $\mathbf{h}_{k,NLoS} \sim \mathcal{CN}(\mathbf{0},\mathbf{R})$. As a result, $\mathbb{E}\{\bm{\mathsf{a}}_k^H\mathbf{R}_C\bm{\mathsf{a}}_k\}=\bm{\mathsf{a}}_k^H\mathbf{R}_C\bm{\mathsf{a}}_k$, $\mathbb{E}\{\mathsf{b}^2\mathbf{h}_{k,NLoS}^H\mathbf{R}_C\mathbf{h}_{k,NLoS}\}=\mathsf{b}^2Tr(\mathbf{R}\mathbf{R}_C)$, $\mathbb{E}\{\mathsf{b} \mathbf{h}_{k,NLoS}^H\mathbf{R}_C\bm{\mathsf{a}}_k\}=0$, and $\mathbb{E}\{\mathsf{b}\bm{\mathsf{a}}_k^H \mathbf{R}_C\mathbf{h}_{k,NLoS}\}=0$. Similarly, the second moment is obtained as
\begin{align} \label{AA3}
n_Z = &\mathbb{E}\{Z^2\} \nonumber\\
=&(\bm{\mathsf{a}}_k^H\mathbf{R}_C\bm{\mathsf{a}}_k)^2+ \mathsf{b}^4[Tr((\mathbf{R}\mathbf{R}_C)^2)+(Tr(\mathbf{R}\mathbf{R}_C))^2]\nonumber\\
&+  2\mathsf{b}^2\bm{\mathsf{a}}_k^H\mathbf{R}_C\mathbf{R}\mathbf{R}_C\bm{\mathsf{a}}_k+ 2\mathsf{b}^2\bm{\mathsf{a}}_k^H\mathbf{R}_C\bm{\mathsf{a}}_k Tr(\mathbf{R}\mathbf{R}_C).
\end{align} 
\noindent Furthermore, the variance of $Z$ is obtained as
\begin{align} \label{AA4}
v_Z =& n_Z-m_Z^2\nonumber\\
=&\mathsf{b}^4Tr((\mathbf{R}\mathbf{R}_C)^2) + 2\mathsf{b}^2\bm{\mathsf{a}}_k^H\mathbf{R}_C\mathbf{R}\mathbf{R}_C\bm{\mathsf{a}}_k.
\end{align} 
We now can match $Z$ with a gamma distribution, i.e., $Z \sim Gamma(\vartheta_k,\varkappa_k)$, where the shape $\vartheta_k = m_Z^2/v_Z$ and the scale $\varkappa_k=v_Z/m_Z$. The result in \eqref{O3} is thus followed.

\section{Outage Probability via a SPA Technique}\label{Appendix_B}
At first, recall that $\mathbf{h}_k=\bm{\mathsf{a}}_k+\mathsf{b} \mathbf{h}_{k,NLoS}$, where $\bm{\mathsf{a}}_k$ is a deterministic vector representing the LoS component, and $\mathbf{h}_{k,NLoS} \sim \mathcal{CN}(\mathbf{0},\mathbf{R})$. Thus, we have $\mathbf{h}_{k} \sim \mathcal{CN}(\bm{\mathsf{a}}_k,\mathsf{b}^2\mathbf{R})$, and $Z = ||\mathbf{g}_k||^2 =\mathbf{h}_k^H\mathbf{R}_C\mathbf{h}_k$ has a quadratic form under correlated non-central complex Gaussian input. To analyze the CDF of $Z$ via a SPA method, we need to transform it into an uncorrelated case. To proceed, we perform a Cholesky factorization such that $\mathbf{R}=\mathbf{F}_0\mathbf{F}_0^H$, where $\mathbf{F}_0$ is a lower triangular matrix \cite{Hor}. Also, let us define $\mathbf{F}=\mathsf{b}\mathbf{F}_0$ for notational convenience. Then, $\mathbf{h}_k$ can be expressed as $\mathbf{h}_k=\bm{\mathsf{a}}_k+\mathbf{F}\bm{\mathsf{q}}_k$, where $\bm{\mathsf{q}}_k \sim \mathcal{CN}(\mathbf{0},\mathbf{I})$. After some algebraic manipulations, we can rewrite the value of $Z$ as
\begin{align} \label{AB6}
Z=&(\bm{\mathsf{a}}_k+\mathbf{F}\bm{\mathsf{q}}_k)^H\mathbf{R}_C(\bm{\mathsf{a}}_k+\mathbf{F}\bm{\mathsf{q}}_k) \nonumber\\
=&(\mathbf{F}^{-1}\bm{\mathsf{a}}_k+\bm{\mathsf{q}}_k)^H\mathbf{F}^H\mathbf{R}_C\mathbf{F}(\mathbf{F}^{-1}\bm{\mathsf{a}}_k+\bm{\mathsf{q}}_k).
\end{align}
\noindent Since $\mathbf{F}^H\mathbf{R}_C\mathbf{F}$ is Hermitian, it can be diagonalized by a unitary matrix $\mathbf{U}$ as $\mathbf{F}^H\mathbf{R}_C\mathbf{F}=\mathbf{U}\mathbf{\Lambda}\mathbf{U}^H$, where $\bm{\Lambda}=\operatorname{diag}(\zeta_1,\zeta_2,\cdots,\zeta_I)$ is a diagonal matrix containing eigenvalues $\zeta_i$ \cite{Hor}. Moreover, let us define $\bm{\tilde{\mathsf{a}}}_k=\mathbf{U}^H\mathbf{F}^{-1}\bm{\mathsf{a}}_k$ and $\bm{\tilde{\mathsf{q}}}_k=\mathbf{U}^H\bm{\mathsf{q}}_k \sim \mathcal{CN}(\mathbf{0},\mathbf{I})$. Then, we can express $Z$ in a standard quadratic form, i.e.,
\begin{align} \label{AB7}
Z=(\bm{\tilde{\mathsf{a}}}_k+\bm{\tilde{\mathsf{q}}}_k)^H\Lambda(\bm{\tilde{\mathsf{a}}}_k+\bm{\tilde{\mathsf{q}}}_k),
\end{align}
\noindent which can be further expressed as
\begin{align} \label{AB8}
Z=\sum_{i=1}^I\zeta_i|\tilde{\mathsf{a}}_k+\tilde{\mathsf{q}}_k|^2=\sum_{i=1}^I\zeta_i Z_i,
\end{align} 
\noindent where $Z_i = |\tilde{\mathsf{a}}_k+\tilde{\mathsf{q}}_k|^2 \sim \chi_2^2(\mu_i)$ is a non-central chi-squared random variable with two degrees of freedom and a non-centrality parameter of $\mu_i=|\mathbf{u}_i^H\mathbf{F}^{-1}\bm{\mathsf{a}}_k|^2$. It is worth noting that $Z$ is a sum of independent variables. Thus, the moment generation function (MGF) of $Z$ can be evaluated as \cite{Sim}
\begin{align} \label{AB9}
\mathcal{M}_g(s)&=\mathbb{E}\{e^{sZ}\}=\prod_{i=1}^I\mathbb{E}\{e^{s\zeta_i Z_i}\} \nonumber\\
&=\prod_{i=1}^I \frac{1}{1-s\zeta_i}\exp\left(\frac{s\zeta_i\mu_i}{1-s\zeta_i}\right), s\zeta_i <1,
\end{align} 
\noindent where the last step is obtained since $\mathbb{E}\{e^{s\zeta_i Z_i}\}=\frac{1}{1-s\zeta_i}\exp\left(\frac{s\zeta_i\mu_i}{1-s\zeta_i}\right), s\zeta_i <1$ \cite{Tzi}. Based on the MGF, we can compute the cumulant generating function (CGF) as
\begin{align} \label{AB10}
\mathcal{C}_g(s)=\log \mathcal{M}_g(s)=\sum_{i=1}^I\left[-\log(1\!-\!s\zeta_i)\!+\!\frac{s\zeta_i\mu_i}{1\!-\!s\zeta_i}\right]\!.
\end{align} 
\noindent Thus, the first and second derivatives of the CGF are obtained, respectively, as
\begin{align} \label{AB11}
\mathcal{C}_g^{'}(s)=\sum_{i=1}^I\left[\frac{\zeta_i}{1-s\zeta_i}+\frac{\zeta_i\mu_i}{(1-s\zeta_i)^2}\right],
\end{align}
\noindent and
\begin{align} \label{AB12}
\mathcal{C}_g^{''}(s)=\sum_{i=1}^I\left[\frac{\zeta_i^2}{(1-s\zeta_i)^2}+\frac{2\zeta_i^2\mu_i}{(1-s\zeta_i)^3}\right].
\end{align}

With the existence of $\mathcal{C}_g(s)$, $\mathcal{C}_g^{'}(s)$, and $\mathcal{C}_g^{''}(s)$, we can now use a SPA method to obtain the CDF of $Z$. In particular, we deploy Newton's method to find the saddle point $s^{\star}$ via solving the equation of $\mathcal{C}_g^{'}(s)=\xi_k$. The CDF is then obtained based on the Lugannani–Rice formula as (cf. \cite{Gur})
\begin{align} \label{AB13}
P_{out,k}=\Phi(\omega_k)+\phi(\omega_k)\left(\frac{1}{\omega_k}-\frac{1}{\varpi_k}\right), 
\end{align}
\noindent where $\Phi(\omega_k)=\frac{1}{\sqrt{2\pi}}\int_{-\infty}^{\omega_k} e^{-x^2/2}dx$ and $\phi(\omega_k)=\frac{1}{\sqrt{2\pi}}e^{-\omega_k^2/2}$ are the CDF and PDF of the standard normal distribution. Also, $\omega_k=sign(s^{\star})\sqrt{2(s^{\star}\xi_k-\mathcal{C}_g(s^{\star}))}$, $\varpi_k=s^{\star}\sqrt{\mathcal{C}_g^{''}(s^{\star})}$, where $sign(x)$ is 1 if $x\geq 0$ and 0 if $x<0$. This completes the proof.

\section{System Model Extension to Imperfect CSI}\label{Appendix_C}
In this Appendix, we extend the system model to cases with channel estimation errors. The channel coefficients from the SIM-HAPS to the $k$-th ground user can be expressed as \cite{Ahn}
\begin{align} \label{AC1}
\mathbf{h}_k=\sqrt{1-\epsilon^2}\mathbf{\tilde{h}}_k+\epsilon\mathbf{e}_k,
\end{align} 
\noindent where $\mathbf{\tilde{h}}_k$ is the estimated channel coefficients, $\epsilon$ denotes the accuracy of the channel estimation (i.e., $\epsilon=0$ indicates no CSI errors), and $\mathbf{e}_k \sim \mathcal{CN}(0,\sigma_{e,k}^2)$ is the channel estimation error with variance $\sigma_{e,k}^2 = \beta_k$. Thus, the received signal at the $k$-th user can be expressed as (cf. (3))
\begin{align} \label{AC2}
y_k=&(\sqrt{1-\epsilon^2}\mathbf{\tilde{h}}_k+\epsilon\mathbf{e}_k)^H\mathbf{B}\boldsymbol{\Psi}_1\mathbf{\bar{w}}_0\sqrt{p_0}s_0+\nonumber\\
&\sum_{r=1}^K(\sqrt{1-\epsilon^2}\mathbf{\tilde{h}}_k+\epsilon\mathbf{e}_k)^H\mathbf{B}\boldsymbol{\Psi}_1\mathbf{\bar{w}}_r\sqrt{p_r}s_r+n_k\nonumber\\ 
=&\sqrt{1-\epsilon^2}\mathbf{\tilde{g}}_k\mathbf{\bar{w}}_0\sqrt{p_0}s_0+\epsilon\mathbf{e}_k^H\mathbf{B}\boldsymbol{\Psi}_1\mathbf{\bar{w}}_0\sqrt{p_0}s_0+ \nonumber\\
& \sum_{r=1}^K\left[\sqrt{1-\epsilon^2}\mathbf{\tilde{g}}_k\mathbf{\bar{w}}_r\sqrt{p_r}s_r+ \epsilon\mathbf{e}_k^H\mathbf{B}\boldsymbol{\Psi}_1\mathbf{\bar{w}}_r\sqrt{p_r}s_r\right]+n_k,
\end{align} 
\noindent where $\mathbf{\tilde{g}}_k \triangleq \mathbf{\tilde{h}}_k^H\mathbf{B}\boldsymbol{\Psi}_1$. The SINRs of the multicast and unicast stream, respectively, become (cf. (4)-(7))
\begin{align} \label{AC3}
\gamma_k^{\mathcal{M}}&=\frac{(1-\epsilon^2)|\mathbf{\tilde{g}}_k\mathbf{\bar{w}}_0|^2p_0}{(1\!-\!\epsilon^2)\!\sum\limits_{r=1}^K\!|\mathbf{\tilde{g}}_k\mathbf{\bar{w}}_r|^2p_r\!+\!\epsilon^2\sigma_{e,k}^2\!\sum\limits_{r=0}^K\!||\mathbf{B}\boldsymbol{\Psi}_1\mathbf{\bar{w}}_r||^2p_r\!+\!N_0} \nonumber\\
&=\frac{(1-\epsilon^2)||\mathbf{\tilde{g}}_k||^2p_0}{(1-\epsilon^2)||\mathbf{\tilde{g}}_k||^2p_k+ \epsilon^2\sigma_{e,k}^2\sum_{r=0}^K||\mathbf{B}\boldsymbol{\Psi}_1\mathbf{\bar{w}}_r||^2p_r+N_0},
\end{align} 
\noindent and
\begin{align} \label{AC4}
\gamma_k^{\mathcal{U}}&=\frac{(1-\epsilon^2)|\mathbf{\tilde{g}}_k\mathbf{\bar{w}}_k|^2p_k}{(1\!-\!\epsilon^2)\!\sum\limits_{\substack{r=1 \\ r\neq k}}^K\!|\mathbf{\tilde{g}}_k\mathbf{\bar{w}}_r|^2p_r\!+\!\epsilon^2\sigma_{e,k}^2\sum\limits_{r=0}^K||\mathbf{B}\boldsymbol{\Psi}_1\mathbf{\bar{w}}_r||^2p_r\!+\! N_0} \nonumber\\
&=\frac{(1-\epsilon^2)||\mathbf{\tilde{g}}_k||^2p_k}{\epsilon^2\sigma_{e,k}^2\sum_{r=0}^K||\mathbf{B}\boldsymbol{\Psi}_1\mathbf{\bar{w}}_r||^2p_r+N_0},
\end{align} 
\noindent The achievable rates of the private messages and the multicast message are then obtained based on (8) and (9), respectively. 

Regarding the problem of optimizing the transmit power $P$ for the maximal EE in the presence of CSI errors, we note that the unicast rate
constraint of (17c) and the multicast rate constraint of (17d) in this case are different from those in perfect CSI. Specifically, based on the SINR defined in \eqref{AC3}-\eqref{AC4}, we have (cf. (19)-(20)) 
\begin{align} \label{AC5}
&P^{unicast}_{min} = (K+1)\times \nonumber\\
&\max_{k}\frac{(2^{R_k^{th}}-1)N_0}{(1-\epsilon^2)||\mathbf{\tilde{g}}_k||^2-\epsilon^2\sigma_{e,k}^2(2^{R_k^{th}}-1)\sum\limits_{r=0}^K||\mathbf{B}\boldsymbol{\Psi}_1\mathbf{\bar{w}}_r||^2},
\end{align}
\noindent and
\begin{align} \label{AC6}
&P^{multicast}_{min} =(K+1)\times \nonumber\\
&\max_{k}\frac{(2^{R_0^{th}}-1)N_0}{(1\!-\!\epsilon^2)||\mathbf{\tilde{g}}_k||^2(2\!-\!2^{R_0^{th}}\!)\!-\!\epsilon^2\sigma_{e,k}^2(2^{R_0^{th}}\!-\!1)\!\!\sum\limits_{r=0}^K\!\!||\mathbf{B}\boldsymbol{\Psi}_1\mathbf{\bar{w}}_r||^2}.
\end{align}
\noindent The remain steps are similar to those in Section IV.B.


\begin{thebibliography}{60}
\bibitem{IMT} “Framework and overall objectives of the future development of IMT for 2030 and beyond,” \textit{ITU-R Recommendation M.2160-0}, Nov. 2023.
\bibitem{Dan} S. Dang, O. Amin, B. Shihada and M.-S. Alouini, ``What should 6G be?", \textit{Nature Electronics}, vol. 3, pp. 20-29, Jan. 2020.
\bibitem{Di} M. Di Renzo, A. Zappone, M. Debbah, M.-S. Alouini, C. Yuen, J. de Rosny, and S. Tretyakov, “Smart radio environments empowered by
reconfigurable intelligent surfaces: How it works, state of research, and the road ahead,” \textit{IEEE J. Sel. Areas Commun.}, vol. 38, no. 11, pp. 2450-2525, 2020.
\bibitem{Wu} Q. Wu, \textit{et al.}, ``Intelligent surfaces empowered wireless network: Recent advances and the road to 6G," \textit{Proc. of the IEEE}, vol. 112, no. 7, pp. 724-763, Jul. 2024.
\bibitem{Aza} M. M. Azari et al., "Evolution of non-terrestrial networks from 5G to 6G: A survey," \textit{IEEE Communications Surveys $\&$ Tutorials}, vol. 24, no. 4, pp. 2633-2672, 2022.
\bibitem{GPP} 3GPP, “Solutions for NR to support non-terrestrial networks (NTN),” \textit{3GPP TR 38.821 V17.0.0}, Mar. 2022 
\bibitem{Ye} J. Ye, J. Qiao, A. Kammoun and M. -S. Alouini, "Non-terrestrial communications assisted by reconfigurable intelligent surfaces,"\textit{Proc. the IEEE}, vol. 110, no. 9, pp. 1423-1465, Sept. 2022.

\bibitem{Kar} G. Karabulut Kurt et al., "A Vision and framework for the high altitude platform station (HAPS) networks of the future," \textit{IEEE Communications Surveys $\&$ Tutorials}, vol. 23, no. 2, pp. 729-779, 2021.
\bibitem{Alf} S. Alfattani, A. Yadav, H. Yanikomeroglu and A. Yongaçoglu, "Resource-efficient HAPS-RIS enabled beyond-cell communications," \textit{IEEE Wireless Commun. Lett.}, vol. 12, no. 4, pp. 679-683, Apr. 2023.
\bibitem{Ji} P. Ji, L. Jiang, C. He, Z. Lian and D. He, "Active RIS aided NOMA for HAP-MISO systems," \textit{IEEE Wireless Commun. Lett.}, vol. 13, no. 8, pp. 2170-2174, Aug. 2024.
\bibitem{Sha} P. Shaik, K. Kishore Garg, P. Kumar Singya, V. Bhatia, O. Krejcar, and M.-S. Alouini, ‘‘On performance of integrated satellite HAPS ground communication: Aerial IRS node vs terrestrial IRS node,’’\textit{ IEEE Open J. Commun. Soc.}, vol. 5, pp. 3775-3791, Jun. 2024.
\bibitem{Azi} A. Azizi and A. Farhang, "RIS meets aerodynamic HAPS: A multi-objective optimization approach," \textit{IEEE Wireless Commun. Lett.}, vol. 12, no. 11, pp. 1851-1855, Nov. 2023.
\bibitem{Gao1} N. Gao, S. Jin, X. Li and M. Matthaiou, "Aerial RIS-assisted high altitude platform communications," \textit{IEEE Wireless Commun. Lett.}, vol. 10, no. 10, pp. 2096-2100, Oct. 2021.

\bibitem{An} J. An et al., ``Stacked intelligent metasurfaces for multiuser beamforming in the wave domain," in \textit{Proc. IEEE Int. Conf. Commun. (ICC)}, pp. 2834-2839, Oct. 2023.
\bibitem{An3} J. An et al., ``Stacked intelligent metasurfaces for multiuser downlink beamforming in the wave domain," \textit{IEEE Trans. Wireless Commun.}, vol. 24, no. 7, pp. 5525-5538, Jul. 2025.
\bibitem{An1} J. An, C. Xu, D. W. K. Ng, G. C. Alexandropoulos, C. Huang, C. Yuen, and L. Hanzo, ``Stacked intelligent metasurfaces for efficient holographic mimo communications in 6G," \textit{IEEE J. Sel. Areas Commun.}, vol. 41, no. 8, pp. 2380-2396, Aug. 2023.
\bibitem{An2} J. An et al., “Stacked intelligent metasurface-aided MIMO transceiver design,” \textit{IEEE Wireless Commun.}, vol. 31, no. 4, pp. 123-131, 2024.
\bibitem{Pap1} A. Papazafeiropoulos, P. Kourtessis, S. Chatzinotas, D. I. Kaklamani, and I. S. Venieris, “Achievable rate optimization for large stacked intelligent metasurfaces based on statistical CSI,” \textit{IEEE Wireless Commun. Lett.}, vol. 13, no. 9, pp. 2337-2341, 2024.
\bibitem{Liu} H. Liu, J. An, G. C. Alexandropoulos, D. W. K. Ng, C. Yuen and L. Gan, "Multi-user MISO with stacked intelligent metasurfaces: A DRL-based sum-rate optimization approach," \textit{IEEE Trans. Cognitive Commun. Net.}, vol. 12, pp. 251-266, 2026.
\bibitem{Per} N. S. Perovi´c, E. E. Bahingayi, and L.-N. Tran, “Energy-efficient designs for SIM-based broadcast MIMO systems,” \textit{IEEE Trans. Commun.}, vol. 72, no. 12, pp. 15881-15894, Dec. 2025.
\bibitem{Niu7} H. Niu et al.,  “Introducing meta-fiber into stacked intelligent metasurfaces for MIMO communications: A low-complexity design with only two layers,” \textit{IEEE Trans. Wireless Commun.}, vol. 25, pp. 3016-3032, 2026.
\bibitem{Yao} X. Yao, J. An, L. Gan, M. Di Renzo, and C. Yuen, “Channel estimation for stacked intelligent metasurface-assisted wireless networks,” \textit{IEEE Wireless Commun. Lett.}, vol. 13, pp. 1349-1353, May 2024.
\bibitem{Pap2} A. Papazafeiropoulos, P. Kourtessis, D. I. Kaklamani and I. S. Venieris, "Channel estimation for stacked intelligent metasurfaces in Rician fading channels," \textit{IEEE Wireless Commun. Lett.}, vol. 14, no. 5, pp. 1411-1415, May 2025.
\bibitem{Jia}  Y. Jia, H. Lu, Z. Fan, B. Wu, F. Qu, M.-J. Zhao, C. Qian, and H. Chen, “High-efficiency transmissive tunable metasurfaces for binary cascaded diffractive layers,” \textit{IEEE Trans. Antennas Propag.}, vol. 72, no. 5, pp. 4532-4540, May 2024.

\bibitem{Hu} Y. Hu, J. Zhang, E. Shi, Y. Lu, J. An, C. Yuen, and B. Ai, ``Joint beamforming and power allocation design for stacked intelligent metasurfaces-aided cell-free massive MIMO systems," \textit{IEEE Trans. Veh. Technol.}, vol. 74, no. 3, pp. 5235-5240, 2025.
\bibitem{Shi} E. Shi, J. Zhang, Y. Zhu, J. An, C. Yuen, and B. Ai, “Uplink performance of stacked intelligent metasurface-enhanced cell-free massive MIMO systems,” \textit{IEEE Trans. Wireless Commun.}, vol. 24, no. 5, pp. 3731-3746, May 2025.
\bibitem{Li} Q. Li, M. El-Hajjar, C. Xu, J. An, C. Yuen and L. Hanzo, "Stacked intelligent metasurface-based transceiver design for near-field wideband systems," \textit{IEEE Trans. Commun.}, vol. 73, no. 9, pp. 3731-3746, Sep. 2025.
\bibitem{Lin} S. Lin et al., ``Stacked intelligent metasurface enabled LEO satellite communications relying on statistical CSI," \textit{IEEE Wireless Commun. Lett.}, vol. 13, no. 5, pp. 1295-1299, May. 2024.
\bibitem{Niu1} H. Niu, J. An, A. Papazafeiropoulos, L. Gan, S. Chatzinotas, and M. Debbah, ``Stacked intelligent metasurfaces for integrated sensing and communications," \textit{IEEE Wireless Commun. Lett.}, vol. 13, no. 10, pp. 2807-2811, 2024.
\bibitem{Lis} S. Li, F. Zhang, T. Mao, R. Na, Z. Wang, and G. K. Karagiannidis, “Transmit beamforming design for ISAC with stacked intelligent metasurfaces,” \textit{IEEE Trans. Veh. Technol.}, vol. 74, pp. 6767-6772, Apr. 2025.
\bibitem{Hua} G. Huang, J. An, Z. Yang, L. Gan, M. Bennis and M. Debbah, “Stacked intelligent metasurfaces for task-oriented semantic communications,” \textit{IEEE Wireless Commun. Lett.}, vol. 14, no. 2, pp. 310-314, Feb. 2025.

\bibitem{GPP1} 3GPP, Architectural enhancements for 5G multicast-broadcast services; Stage 2 (Release 19), 3GPP TS 23.247, V19.2.0, Tech. Spec. Group Services and System Aspects, Jun. 2025.
\bibitem{Lin} X. Lin, Y. Rivenson, N. T. Yardimci, M. Veli, Y. Luo, M. Jarrahi, and A. Ozcan, “All-optical machine learning using diffractive deep neural networks,” \textit{Science}, vol. 361, no. 6406, pp. 1004-1008, Jul. 2018.
\bibitem{Kim} D. Kim, F. Khan, C. V. Rensburg, Z. Pi, and S. Yoon, “Superposition of broadcast and unicast in wireless cellular systems,” \textit{IEEE Commun. Mag.}, vol. 46, no. 7, pp. 110-117, Jul. 2008.
\bibitem{Liz} Z. Li, S. Wang, S. Han, and C. Li, “Non-orthogonal broadcast and unicast joint transmission for multibeam satellite system,” \textit{IEEE Trans. Broadcast.}, vol. 69, no. 3, pp. 647-660, Sep. 2023.
\bibitem{Vu} T. -H. Vu, Q. -V. Pham, D. B. da Costa and S. Kim, "Rate-splitting multiple access-assisted THz-based short-packet communications," \textit{IEEE Wireless Commun. Lett.}, vol. 12, pp. 2218-2222, Dec. 2023.

\bibitem{Gur} S. Guruacharya, H. Tabassum and E. Hossain, ``Saddle point approximation for outage probability using cumulant generating functions," \textit{IEEE Wireless Commun. Lett.}, vol. 5, no. 2, pp. 192-195, Apr. 2016.
\bibitem{LR} R. Lugannani and S. O. Rice, “Saddle point approximation for the distribution of the sum of independent random variables,” \textit{Adv. Appl. Probab.}, vol. 12, no. 2, pp. 475-490, Jun. 1980.

\bibitem{Esm}  Z. Esmaeilbeig, A. Eamaz, K. V. Mishra, and M. Soltanalian, "Joint waveform and passive beamformer design in multi-IRS-aided radar", in \textit{Proc. IEEE International Conference on Acoustics, Speech and Signal Processing (ICASSP)}, 2023.

\bibitem{Sca} L. E. Scales, \emph{Introduction to Non-Linear Optimization}. Macmillan Publisher, 1985.

\bibitem{Gao} J. Gao, C. Zhong, X. Chen, H. Lin, and Z. Zhang, “Unsupervised learning for passive beamforming,” \textit{IEEE Commun. Lett.}, vol. 24, no. 5, pp. 1052-1056, 2020.
\bibitem{Mol} P. Molchanov, S. Tyree, T. Karras, T. Aila, and J. Kautz, "Pruning convolutional neural networks for resource efficient inference," in \textit{Proc. Int. Conf. Learning Representations} (ICLR), 2017.
\bibitem{Ji5} P. Ji, L. Jiang, C. He, Z. Lian and D. He, "Energy-efficient beamforming for beamspace HAP-NOMA systems," \textit{IEEE Commun. Letts.}, vol. 25, no. 5, pp. 1678-1681, May 2021.
\bibitem{Ahn} K. S. Ahn and R. W. Heath, "Performance analysis of maximum ratio combining with imperfect channel estimation in the presence of cochannel interferences," \textit{IEEE Trans. Wireless Commun.}, vol. 8, no. 3, pp. 1080-1085, Mar. 2009.
\bibitem{Hor} R. A. Horn and C. R. Johnson, \textit{Matrix Analysis}, 2nd ed., Cambridge University Press, 2013.
\bibitem{Sim} M. K. Simon and M.-S. Alouini, \textit{Digital Communication over Fading Channels}, 2nd ed. Hoboken, NJ, USA: Wiley, 2005.
\bibitem{Tzi} G. Tziritas, ``On the distribution of positive-definite Gaussian quadratic forms," \textit{IEEE Trans. Inf. Theory}, vol. 33, no. 6, pp. 895-906, Nov. 1987.

\end{thebibliography}
\end{document}